\documentclass[numreferences]{kluwer}

\usepackage{epsfig,graphics,amssymb,subeqnarray,verbatim,color,subfig,sidecap}
\usepackage{color}
\usepackage{comment}
\usepackage{bm}

\def\SS{{\mathbf{S}}}

\def\P{{\mathbf{P}}}

\def\F{{\mathbf{F}}}
\def\u{{\mathbf{u}}}

\def\U{{\mathbf{U}}}
\def\T{{\mathbf{T}}}
\def\I{{\mathbf{I}}}

\def\v{{\mathbf{v}}}

\def\n{{\mathbf{n}}}

\def\G{{\mathbf{G}}}
\def\R{{\mathbf{R}}}
\def\r{{\mathbf{r}}}

\def\zetaB{{\bm{\zeta}}}
\def\sigmaB{{\bm{\sigma}}}
\def\OmegaB{{\bm{\Omega}}}

\def\alphaB{{\bm{\alpha}}}
\def\betaB{{\bm{\beta}}}
\def\gammaB{{\bm{\gamma}}}

\def\varepsilon{\epsilon}

\def\e{{\mathbf{e}}}

\begin{document}
\begin{article}

\begin{opening}
\title{Generalized squirming motion of a sphere}            

\author{On Shun \surname{Pak}}
\institute{Department of Mechanical and Aerospace
Engineering, Princeton University, 
Olden Street, Princeton, NJ, 08544-5263, USA}          
\author{Eric \surname{Lauga}\email{e.lauga@damtp.cam.ac.uk}} 
\institute{Department of Applied Mathematics and Theoretical Physics, University of Cambridge, 
 Centre for Mathematical Sciences, Wilberforce Road, Cambridge CB3 0WA, UK}                               




\runningtitle{Generalized squirming motion of a sphere}
\runningauthor{On Shun Pak and Eric Lauga}



\begin{abstract} 
A number of swimming microorganisms such as ciliates (\textit{Opalina}) and multicellular colonies of flagellates (\textit{Volvox}) are approximately spherical in shape and swim using beating arrays of cilia or short flagella covering their surfaces. Their physical actuation on the fluid  may be mathematically modeled as the generation of surface velocities on a continuous spherical surface -- a model known in the literature  as squirming, which has been used to address various aspects of the biological physics of locomotion. Previous analyses of squirming assumed axisymmetric fluid motion and hence restricted all swimming kinematics to take place along a line. In this paper we generalize squirming to three spatial dimensions. We derive analytically the flow field surrounding a spherical squirmer with arbitrary surface motion, and use it to derive its three-dimensional translational and rotational swimming kinematics. We then use our results to physically interpret the  flow field induced by the swimmer in terms of fundamental flow singularities  up to terms decaying spatially as $\sim 1/r^3$. Our results will enable to  develop new models in biological physics, in particular in the area of hydrodynamic interactions and  collective locomotion.
\end{abstract}




\end{opening}

\section{Introduction} 
\label{sec:Intro}

Due to their small sizes, microorganisms inhabit a world where viscous forces dominate and inertial effects are negligible. The Reynolds number, which characterizes the relative importance of inertial to viscous forces, ranges typically from $10^{-5}$ for the smallest bacteria up to $10^{-2}$  for spermatozoa \cite{Brennen1977}. Fluid-based locomotion of microorganisms is vital in a number of biological processes, including  reproduction, locating nutrient sources, preying, and escaping from predation \cite{Vogel1996, Bray2000, Guasto2012}. Physically, locomotion at low Reynolds numbers suffers from the constraints due to the absence of inertia, mathematically manifested by the linearity and time-independence of the governing equation -- the Stokes equations \cite{Lauga2009}. Purcell  illustrated the  difficulties encountered in small-scale locomotion by introducing his scallop theorem \cite{Purcell1977}, which states that any reciprocal motion (body deformation possessing a time-reversal symmetry) cannot lead to any net propulsion at zero Reynolds number  \cite{Ishimoto2012}.

\begin{figure}[t]
\begin{center}
\includegraphics[width=0.9\textwidth]{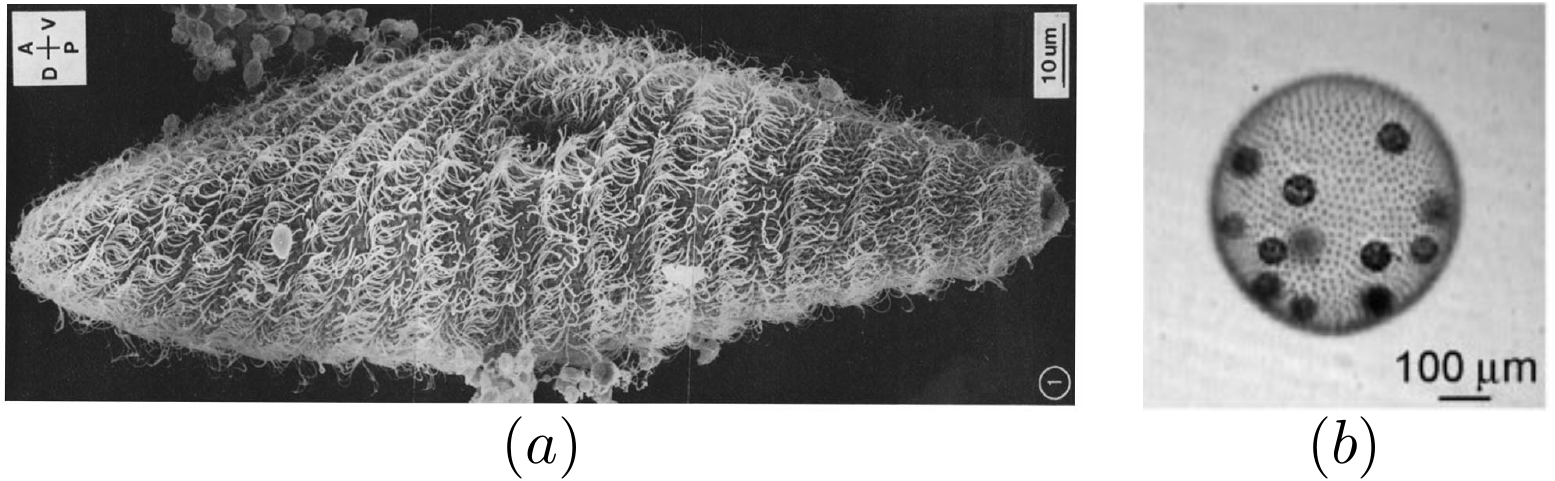}
\caption{(a) Ciliary motion of  \textit{Paramecium} \cite{Tamm1972}, and (b) flagellar motion of \textit{Volvox} \cite{Solari2006}. All images were reprinted with permission: (a) from Tamm \cite{Tamm1972} Copyright \copyright 1972 The Rockefeller University Press. (b) from Solari \textit{et al.} \cite{Solari2006} Copyright \copyright 2006 The National Academy of Sciences of the United States of America.}
\label{fig:Ciliates}
\end{center}
\end{figure}

Nature showcases a variety of mechanisms able to overcome the constraints of the theorem and achieve micro-propulsion. Many cells use one or more appendages, called a flagellum (plural, flagella), for propulsion. Traveling waves are then propagated along the flagellum either by  internal bending (seen, for example, in the spermatozoon of eukaryotic cells) or passive rotation of a rigid helical flagellum (the case of swimming bacteria), in both cases allowing to break the time-reversal symmetry and hence escape from the constraints of the scallop theorem \cite{Lauga2009}.  A number of microorganisms possess  multiple flagella. \textit{Escherichia coli} is a bacterium with a few helical flagella that can wrap into a bundle to move the cell forward when the motor turns in a specific direction. \textit{Chlamydomonas reinhardtii} is an alga (eukaryotic cell) with two flagella. Ciliates such as \textit{Opalina} and \textit{Paramecium} (illustrated in Fig.~\ref{fig:Ciliates}a) and colonies of flagellates such as \textit{Volvox} (shown in Fig.~\ref{fig:Ciliates}b) have their surface covered by arrays of cilia (or short flagella) beating in a coordinated fashion  \cite{Brennen1977}.

Over the past  60 years, theoretical and experimental studies on locomotion of microorganisms have improved our understanding of life under the microscope \cite{Brennen1977, Fauci2006, Lauga2009}. Topics of recent active interest include the locomotion of cells in environments with complex geometries \cite{Galajda2007, DiLeonardo2010, Denissenko2012} and complex fluids \cite{Lauga2007,Fu2007,Shen11, Liu2011}, the role of motility in the formation of biofilms \cite{Pratt1998,Lemon2007,Houry2010, Berke2008}, collective dynamics of active particles \cite{Ramaswamy2010, Koch2011}, and the effect of Brownian noise on swimming \cite{Berg1993, Howse2007, Romanczuk2012}. Considerable attention has also been given on the design of artificial microscopic swimmers \cite{Ebbens2010} for potential biomedical applications such as microsurgery and targeted drug delivery \cite{Nelson2010}. 

Historically, Taylor \cite{Taylor1951} pioneered the theoretical modeling of flagellar hydrodynamics by analyzing the motion of a waving sheet in Stokes flows.  The propulsion speed of the  sheet was solved asymptotically in the limit of small waving amplitude  compared to its  wavelength. Subsequently, most of the  theoretical studies in the field have derived their results  asymptotically, meaning they are physically valid only in specific mathematical limits: small amplitudes \cite{Taylor1951,Reynolds1965,Lauga2007,Fu2007}, long-wavelength \cite{Katz1974,Balmforth2010}, slender filaments \cite{Gray1955,Cox1970,Batchelor1970,Chwang1971,Keller1976,Johnson1980}, or in the far-field  \cite{Pedley1992, Berke2008}. As a result, very few  exact solutions for swimming in Stokes flows exist.

The most popular exact solution is originally due to Lighthill \cite{Lighthill1952} and Blake \cite{Blake1971b} and was  developed to address the propulsion of ciliates. In their model, sometimes referred as the envelope model, the motion of  closely packed cilia tips  are modeled as a continuously deforming surface (envelope) over the  body of the organism, taken to be of  spherical shape  \cite{Lighthill1952, Blake1971b}.  The deformation of the envelope can then be expanded about the surface of the spherical cell body and to leading order, the action of cilia is represented by distributions of radial and tangential velocities on the spherical surface. Lighthill \cite{Lighthill1952}  first derived the exact solution to the Stokes equation due to such a squirming motion on a sphere, with subsequent corrections and generalizations by  Blake \cite{Blake1971b}. Since then, the squirmer model has been all but adopted as the hydrogen atom of low-Reynolds number swimming. Originally developed to specifically model the swimming of ciliates, the squirmer model can also be useful in studying other types of swimming microorganisms, broadly categorized as ``pushers" and ``pullers" \cite{Lauga2009}. Pushers  obtain their thrust from the rear part of their body, such as the swimming of all peritrichous bacteria (like \textit{Escherichia coli}). In contrast,  for pullers the thrust comes from their front part, such as the breaststroke swimming of algae genus \textit{Chlamydomonas}. The squirmer model can represent pushers and pullers by correspondingly changing the surface actuation (squirming profile), rendering it a general model to investigate the locomotion of microorganisms. As such it was used  to study many problems including hydrodynamic interactions of swimmers \cite{Ishikawa2006, Drescher2009}, suspension dynamics \cite{Ishikawa2007, Ishikawa2007b}, nutrient transport and uptake by microorganisms \cite{Magar2003, Magar2005, Michelin2011}, optimal locomotion \cite{Michelin2010}, as well as non-Newtonian \cite{Zhu11, Zhu12} and inertial effects \cite{Wang2012}.

Most studies on squirming motion follow the notation of Lighthill  \cite{Lighthill1952} and Blake \cite{Blake1971b} and assume that the surface distortion is axisymmetric. This simplifies the analysis significantly and results in swimmers undergoing swimming along their axis of symmetry only.  Real microorganisms, however, actuate the fluid in a  non-axisymmetric fashion. The protozoan \textit{Paramecium}, for instance, rotates as it swims and has a helical distribution of cilia (Fig.~\ref{fig:Ciliates}a). Stone \& Samuel \cite{Stone1996} derived formulae which relate the translational and rotational velocities of a squirmer to the arbitrary squirming profiles on the sphere via the reciprocal theorem. In experiments with \textit{V. carteri}, Drescher \textit{et al.} \cite{Drescher2010} measured non-axisymmetric squirming profiles and utilized these reciprocal relations to analyze the swimming kinematics of the cell. However, the reciprocal relations do not give any information on the flow surrounding the squirmer.

In this paper, we generalize the classical squirming results to non-axisymmetric actuation.  Using Lamb's general solution in Stokes flow \cite{Lamb1932}, we derive  analytically the  exact solution for the  flow field surrounding the swimmer, together with the swimming kinematics, for a general non-axisymmetric squirmer. Lamb's general solution is ideally suited for problems with spherical or nearly spherical \cite{Brenner1964} geometries, and  a  detailed description of the solution and its applications can be found in classical textbooks \cite{Happel1973,Kim1991}. Our results will be useful for addressing the role of non-axisymmetric actuation in a variety of problems in the biological physics of locomotion, including  feeding and sensing, and the rheology of active suspensions. Furthermore, from a fundamental fluids perspective, our study allows to make the link between arbitrary surface motion and the appearance of non-axisymmetric flow singularities.

The structure of the paper is the following. The problem is mathematically formulated in Sec.~\ref{sec:Formulation}, followed by a summary of the axisymmetric case in Sec.~\ref{sec:Axis}, where the swimming kinematics (Sec.~\ref{sec:AxisSwimming}) and flow structure (Sec.~\ref{sec:AxisPhysics}) are presented. We then generalize the analysis to a non-axisymmetric squirmer in Sec.~\ref{sec:NonAxisMain}, where we present the swimming kinematics (Sec.~\ref{sec:NonAxis} \& Sec.~\ref{sec:reciprocal}), the three-dimensional flow structure (Sec.~\ref{sec:NonAxisPhysics}), the rate of work due to swimming (Sec.~\ref{sec:Rate}), and the decomposition of arbitrary surface velocities in the form of Lamb's general solution (Sec.~\ref{sec:Arbitrary}). We  conclude in Sec.~\ref{sec:Discuss}. In Appendix \ref{app.singularities}, we include detailed expressions of flow singularities used throughout the  paper. While in the main text we present the results of a squirmer with purely tangential deformation, we include the more general case of  non-axisymmetric squirmer with  radial deformation in Appendices \ref{app:AppendixSwimming}--\ref{app:AppendixGeneral}.

\section{Formulation}\label{sec:Formulation}

\begin{figure}[t]
\centering
\includegraphics[width=0.5\textwidth]{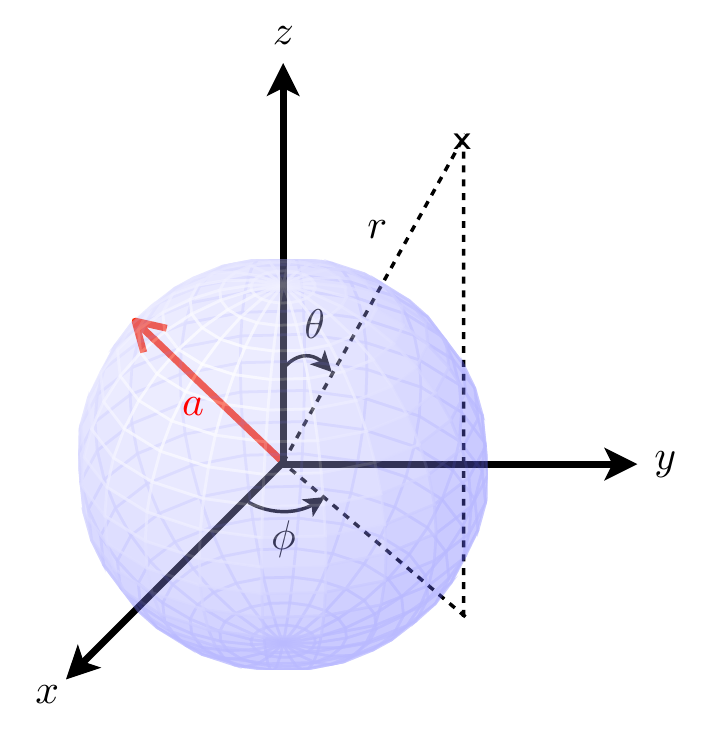}
\caption{Geometrical setup of a spherical squirmer of radius $a$. Using   spherical coordinates, we note the radial coordinate, the polar, and azimuthal angles  as $r$, $\theta$, and $\phi$ respectively.}
\label{coordinates}
\end{figure}

{We model theoretically the motion of a spherical ciliate of radius $a$ in an incompressible fluid at zero Reynolds number using spherical coordinates (Fig.~\ref{coordinates}), with $\e_r$, $\e_\theta$, and $\e_\phi$ as the basis vectors. Following the envelope model, the action of the cilia is represented by a general squirming profile (tangential and radial surface velocities) over the spherical surface at $r=a$. The fluid around the  squirmer is governed by the Stokes equation
\begin{eqnarray}
\eta \nabla^2 \u &=& \nabla p, \label{eqn:Stokes}
\end{eqnarray}
and the continuity equation for incompressible flows
\begin{eqnarray}
\nabla \cdot \u &=& 0,\label{eqn:Continuity}
\end{eqnarray}
where $\u = u_r \e_r + u_\theta \e_\theta + u_\phi \e_\phi$ and $p$ represent the velocity and pressure fields respectively, and $\eta$ denotes the dynamic viscosity of the fluid. In the main text, we present only the results in the case of purely tangential squirming motion (\textit{i.e.~}no radial surface velocities), $u_r(r=a, \theta, \phi) = 0$, as it is the most widely used squirming model in the literature. The more general case which includes the radial deformation is presented in Appendices \ref{app:AppendixSwimming}--\ref{app:AppendixWork}  for completeness.}

A general solution to the Stokes equation can be obtained by constructing the homogeneous ($\u_H$) and particular ($\u_P$) solutions. The homogenous solution can be constructed as \cite{Lamb1932, Happel1973, Kim1991}
\begin{eqnarray}
\u_H = \nabla \Phi + \nabla \times (\r \chi),
\end{eqnarray}
where $\r$ is the position vector, and $\Phi$ and $\chi$ are both harmonic functions
\begin{eqnarray}
\nabla^2 \Phi = \nabla^2 \chi = 0.
\end{eqnarray}
One can  expand the functions $\Phi$ and $\chi$ in series of solid spherical harmonics, $\Phi = \sum_{n=-\infty}^\infty \Phi_n$ and $\chi = \sum_{n=-\infty}^\infty \chi_n$, where $\Phi_n$ and $\chi_n$ denote spherical harmonics of order $n$  as
\begin{eqnarray}
\Phi_n &=& r^n \sum_{m=0}^n P_n^m(\mu) (b_{mn} \cos m\phi +\tilde{b}_{mn} \sin m\phi),\\
\chi_n &=& r^n \sum_{m=0}^n P_n^m(\mu) (c_{mn} \cos m\phi +\tilde{c}_{mn} \sin m\phi).
\end{eqnarray}
Here we have denoted  $\mu = \cos\theta$ and $P_n^m (\mu)$ are the associated Legendre polynomials \cite{Abramowitz1965, Jeffrey2007} of order $m$ and degree $n$, defined as solutions to the linear differential equation for $f(\mu)$ 
\begin{eqnarray}
\frac{d}{d\mu} \left[ (1-\mu^2) \frac{df}{d\mu} \right] + \left[ n(n+1) - \frac{m^2}{1-\mu^2} \right] f = 0.
\end{eqnarray}

By taking the divergence of Eq.~\ref{eqn:Stokes} and utilizing the continuity equation, Eq.~\ref{eqn:Continuity}, we obtain that the pressure  satisfies the Laplace equation $\nabla^2 p =0$. The pressure  is therefore also harmonic, and we can again expand it in a series of solid spherical harmonics, $p = \sum_{n=-\infty}^{\infty} p_n$, where 
\begin{eqnarray}
p_n = r^n \sum_{m=0}^n P_n^m(\mu)(a_{mn} \cos m\phi + \tilde{a}_{mn} \sin m \phi) .
\end{eqnarray}
One can then use these results to construct a particular solution to the Stokes equation as
\begin{eqnarray}
\u_P = \sum_{n=-\infty}^{\infty} \left[ \frac{(n+3)r^2 \nabla p_n}{2\eta (n+1)(2n+3)} - \frac{n\r p_n}{\eta (n+1)(2n+3)}  \right],
\end{eqnarray}
where $r = |\r|$ denotes the magnitude of the position vector.

Superimposing the homogenous and particular solutions gives the general solution
\begin{eqnarray}
\u =&& \sum_{n=-\infty}^\infty  \bigg[ \frac{(n+3)r^2 \nabla p_n}{2\eta (n+1)(2n+3)} - \frac{n\r p_n}{\eta (n+1)(2n+3)}    \nonumber \\ 
&+& \nabla\Phi_n+ \nabla \times (\r \chi_n) \bigg], \label{eqn:Lamb}
\end{eqnarray}
usually named after Lamb \cite{Lamb1932}.

Here we require that the solution decays at infinity ($r \rightarrow \infty$) and hence all harmonics of positive order are discarded.
In addition, following Brenner \cite{Brenner1964}, we replace $n$ by $-(n+1)$ in Eq.~\ref{eqn:Lamb} to obtain the form of Lamb's solution convenient for exterior problems
\begin{eqnarray}\label{firstgen}
\u (r, \theta, \phi) =&& \sum_{n=1}^{\infty} \bigg[- \frac{(n-2) r^2 \nabla p_{-n-1}}{2 \eta n (2n-1)} + \frac{(n+1) \r p_{-n-1}}{\mu n(2n-1)} \nonumber  \\
&+& \nabla \Phi_{-n-1} + \nabla \times (\r \chi_{-n-1})  \bigg],
\end{eqnarray}
where
\begin{eqnarray}
p_{-n-1} &=& r^{-n-1} \sum_{m=0}^{n} P^m_n (A_{mn} \cos m \phi + \tilde{A}_{mn} \sin m\phi), \label{eqn:p}\\
\Phi_{-n-1} &=& r^{-n-1} \sum_{m=0}^{n} P^m_n (B_{mn} \cos m \phi + \tilde{B}_{mn} \sin m\phi), \label{eqn:phi}\\
\chi_{-n-1} &=& r^{-n-1} \sum_{m=0}^{n} P^m_n (C_{mn} \cos m \phi + \tilde{C}_{mn} \sin m\phi), \label{eqn:chi}
\end{eqnarray}
and the pressure field is given by $p = \sum_{n=1}^\infty  p_{-n-1}$.
Notice that the solutions of the case $n=0$ have also been discarded since they correspond to sources and sinks, which are unphysical in problems related to rigid particles \cite{Lighthill1952, Blake1971b,Kim1991}.

After performing all the differential operations in Eq.~\ref{firstgen}, Lamb's general solution in spherical coordinates $\u = u_r \e_r + u_\theta \e_\theta + u_\phi \e_\phi$ takes the form
\begin{eqnarray}
u_r   =&&\sum_{n=1}^\infty \sum_{m=0}^n  \frac{(n+1) P^m_n}{2 (2n-1) \eta r^{n+2}} \bigg\{ \left[ A_{mn} r^2 - 2 B_{mn} (2n-1) \eta \right] \cos m \phi  \nonumber \\
&+& \left( \tilde{A}_{mn} r^2 - 2 \tilde{B}_{mn} (2n-1) \eta  \right) \sin m \phi   \bigg\}, \label{eqn:ur}\\
u_\theta  =&& \sum_{n=1}^\infty \sum_{m=0}^n \frac{1}{2 r^{n} \sin \theta} \bigg \{\sin^2 \theta P_n^{m'} \bigg[  \frac{ n-2}{n(2n-1) \eta} (A_{mn} \cos m \phi + \tilde{A}_{mn} \sin m \phi)  \nonumber \\
&-& \frac{2}{r^2}  (B_{mn} \cos m \phi + \tilde{B}_{mn} \sin m \phi) \bigg]  \nonumber \\
&+& \frac{2m}{r} P^m_n (\tilde{C}_{mn} \cos m\phi - C_{mn} \sin m \phi )  \bigg \}, \label{eqn:utheta}\\
u_\phi =&&\sum_{n=1}^\infty \sum_{m=0}^n \frac{1}{2r^n \sin \theta} \bigg\{ mP_n^m \bigg[ \frac{(n-2)}{n(2n-1) \eta} P^m_n (-\tilde{A}_{mn} \cos m \phi + A_{mn} \sin m\phi) \nonumber \\
&-& \frac{2}{r^2} (-\tilde{B}_{mn} \cos m \phi + B_{mn} \sin m \phi) \bigg]  \nonumber \\
&+& \frac{2}{r} \sin^2\theta P_n^{m'} (C_{mn} \cos m \phi + \tilde{C}_{mn} \sin m \phi) \bigg \} \cdot \label{eqn:uphi}
\end{eqnarray}
Here we have employed a recursion expression of associated Legendre polynomials 
\begin{equation}
(n+1) \mu P_n^m -(1+n-m) P^m_{n+1} = (1-\mu^2) P_n^{m'}
\end{equation}
to simplify the equations \cite{Abramowitz1965, Jeffrey2007} (the primes represent differentiation with respect to the variable $\mu$).

From the radial velocity component, Eq.~\ref{eqn:ur}, the requirement of purely tangential deformation leads to the relations
\begin{equation}
A_{mn} = \frac{2 (2n-1) \eta}{a^2} B_{mn}, \ \ \tilde{A}_{mn} = \frac{2 (2n-1) \eta}{a^2} \tilde{B}_{mn} \label{eqn:purelyTangential}.
\end{equation}
Enforcing the  conditions in Eq.~\ref{eqn:purelyTangential}, the general flow field due to purely tangential squirming motion becomes
\begin{eqnarray}
u_r   =&& \sum_{n=1}^\infty \sum_{m=0}^n \frac{(n+1) P^m_n}{  r^{n+2}} \left( \frac{r^2}{a^2}-1 \right) \left[  B_{mn}\cos m \phi +  \tilde{B}_{mn} \sin m \phi   \right], \label{eqn:urTan}\\
u_\theta  =&& \sum_{n=1}^\infty \sum_{m=0}^n  \bigg [   \sin\theta P^{m'}_n   \left( \frac{n-2}{n a^2 r^n}-\frac{1}{r^{n+2}} \right) \left( B_{mn} \cos m\phi +  \tilde{B}_{mn} \sin m\phi \right) \nonumber \\
&+& \frac{mP^m_n}{r^{n+1} \sin \theta}  (\tilde{C}_{mn} \cos m\phi - C_{mn} \sin m \phi )  \bigg ] \label{eqn:uthetaTan},\\
u_\phi  =&& \sum_{n=1}^\infty \sum_{m=0}^n \bigg[ \frac{\sin \theta P^{m'}_n}{r^{n+1}}  (C_{mn} \cos m \phi + \tilde{C}_{mn} \sin m \phi)  \nonumber \\
&-& \frac{mP^m_n}{\sin \theta} \left( \frac{n-2}{na^2 r^n}-\frac{1}{r^{n+2}}\right) \left(\tilde{B}_{mn} \cos m \phi - B_{mn} \sin m \phi \right) \bigg ]  \label{eqn:uphiTan}.
\end{eqnarray}

To reiterate, the above flow fields decay at infinity in the laboratory frame and correspond to  purely tangential velocities at the body surface;  the case with radial velocities is detailed in Appendices \ref{app:AppendixSwimming}--\ref{app:AppendixWork}.

Note that for simplicity in this paper we consider a neutrally buoyant squirmer, where the buoyancy force from the fluid  balances the gravitational force on the squirmer. Hence, there is no net force and torque acting on the fluid (the force- and torque-free conditions).  Should there be a density offset between the squirmer and the fluid it would result in a net force \cite{Drescher10D} and thus would add  a Stokeslet component (Appendix \ref{app:Stokeslet}) to the flow field around the squirmer, which can  be superimposed to the results of the current work.

\section{Axisymmetric Squirming Motion}\label{sec:Axis}

In this section, we use Lamb's general solution in order to  reproduce the axisymmetric results first derived by Lighthill \cite{Lighthill1952} and Blake \cite{Blake1971b}. The analysis also identifies new axisymmetric modes.  With the general solution, Eqs.~\ref{eqn:urTan}--\ref{eqn:uphiTan}, the axisymmetric flow field ($m=0$) reduces to
\begin{eqnarray}
\u(r, \theta, \phi) =&& \sum_{n=1}^\infty \frac{(n+1)P_n}{r^{n+2}} \left( \frac{r^2}{a^2} -1\right) B_{0n}  \e_r \nonumber \\
&+& \sum_{n=1}^\infty \sin \theta P_n^{'} \left( \frac{n-2}{na^2 r^n}-\frac{1}{r^{n+2}} \right) B_{0n}  \e_\theta \nonumber \\
&+& \sum_{n=1}^\infty \frac{\sin \theta P_n^{m'}}{r^{n+1}} C_{0n}  \e_\phi\label{eqn:flowAxis},
\end{eqnarray}
for purely tangential squirming motion, where we have denoted $P_n(\mu) = P_n^0(\mu)$  the Legendre polynomials of degree $n$. The surface velocities on the sphere have the form
\begin{eqnarray}
\u(a, \theta, \phi) = \sum_{n=1}^\infty  - \frac{2 \sin \theta P^{'}_n }{a^{n+2} n} B_{0n} \e_\theta + \sum_{n=1}^\infty   \frac{\sin \theta P^{'}_n}{a^{n+1}}  C_{0n} \e_\phi. \label{eqn:flowAxisBC}
\end{eqnarray}
Upon setting $C_{0n}=0$ and using a simple rescaling 
\begin{eqnarray}
B_{0n} = - a^{n+2} B_n/(n+1), \label{eqn:rescale}
\end{eqnarray}
the above surface velocities reduce to the form used in Lighthill \cite{Lighthill1952} and Blake \cite{Blake1971b}, where $B_n$ are the coefficients used in their work.

Here we have identified new axisymmetric modes, denoted $C_{0n}$, acting in the azimuthal direction $\phi$ (right hand side of Eq.~\ref{eqn:flowAxisBC}) and the corresponding flow fields (last sum in Eq.~\ref{eqn:flowAxis}), which were not accounted for in previous works. While Stone \& Samuel \cite{Stone1996} and Drescher \textit{et al.} \cite{Drescher2010} employed the reciprocal theorem to discuss the swimming kinematics of a squirmer subject to arbitrary squirming profiles, the current results  complete the analysis of axisymmetric squirming motion by providing the whole flow field. The physical interpretation of these new axisymmetric modes is discussed below in Sec.~\ref{sec:AxisPhysics}.

\subsection{Swimming of an axisymmetric squirmer}\label{sec:AxisSwimming}
When studying the swimming of a squirmer, it is best to think about the problem in two separate steps. In the first step, we consider the above solution, Eq.~\ref{eqn:flowAxis}, with boundary conditions from Eq.~\ref{eqn:flowAxisBC} so that   the squirmer is fixed in space (by an external force) and not allowed to move. This is sometimes referred  to as the ``pumping problem'' in the literature. In the second step, we allow the squirmer to move freely and compute the induced translational ($\U$) and rotational ($\OmegaB$) velocities, given the boundary actuation  in the pumping problem, Eq.~\ref{eqn:flowAxisBC}. This allows the separation of the surface velocities due to the boundary actuation of the squirmer from the contribution due to the induced translation and rotation. To obtain the overall flow field, $\v$,  of a swimming squirmer, we superimpose the solution of the pumping problem, $\u$, to the flow fields due to the induced translation, $\u_T$, and rotation, $\u_R$, and thus write
\begin{eqnarray}
\v = \u + \u_T + \u_R . \label{eqn:PumpSwimg}
\end{eqnarray}

This first step (computing $\u$) was  accomplished in Eq.~\ref{eqn:flowAxis}. We now determine the unknown swimming kinematics, $\{\U, \OmegaB\}$, when the squirmer is free to move. This involves computing all the forces and torques acting on the swimming squirmer: the fluid force, $\F_p$, and torque, $\T_p$, due to the boundary actuation in the pumping problem, and the drag, $\F_s$, and torque, $\T_s$, due to the induced translation and rotation of the squirmer. We then enforce the overall force-free and torque-free conditions  in swimming problems of Stokes flows as
\begin{eqnarray}
\F_p + \F_s &= \mathbf{0}, \label{eqn:forceBalance}\\
\T_p + \T_s &= \mathbf{0} . \label{eqn:torqueBalance}
\end{eqnarray}
The drag and torque due to the translation and rotation of a spherical squirmer are simply {$\F_s = -6 \pi \eta a \U$ and $\T_s = -8 \pi \eta a^3 \OmegaB$} respectively.  For the contributions from the pumping problem, the net force and torque due to the boundary actuation can be conveniently computed in Lamb's general solution as $\F_p= -4 \pi \nabla (r^3 p_{-2})$ and $\T_p = -8 \pi \eta \nabla (r^3 \chi_{-2})$ respectively \cite{Happel1973, Kim1991}. The force and torque balances, Eqs.~\ref{eqn:forceBalance}--\ref{eqn:torqueBalance}, therefore become
\begin{eqnarray}
-4 \pi \nabla (r^3 p_{-2}) - 6 \pi \eta a \U &= \mathbf{0}, \label{eqn:forceBalanceD}\\
-8 \pi \eta \nabla (r^3 \chi_{-2}) -8 \pi \eta a^3 \OmegaB &= \mathbf{0}.\label{eqn:torqueBalanceD}
\end{eqnarray}

The solutions for $p_{-2}$ and $\chi_{-2}$ are given by Eqs.~\ref{eqn:p} and  \ref{eqn:chi} respectively, and we thus obtain the swimming kinematics 
\begin{eqnarray}
\U &=& -\frac{2}{3 \eta a} \nabla \left(r P_1 \right) A_{01} = - \frac{4B_{01}}{3a^3}  \e_z, \label{eqn:AxisSpeed}\\
\OmegaB &=& -\frac{\nabla \left(rP_1\right)}{a^3} C_{01} = - \frac{C_{01}}{a^3} \e_z, \label{eqn:AxisRotation}
\end{eqnarray}
where we have used Eq.~\ref{eqn:purelyTangential}. Note that the propulsion and rotational velocities may also be obtained using a reciprocal theorem approach \cite{Stone1996} as is discussed in Sec.~\ref{sec:reciprocal}. The translational swimming velocity agrees with that given by Lighthill \cite{Lighthill1952} and Blake \cite{Blake1971b} provided  the rescaling from Eq.~\ref{eqn:rescale} is used, giving $\U = 2B_1/3  \e_z$. 

For an axisymmetric squirmer, propulsion and rotation can only occur in the same direction (here, the $z$-direction), and the squirmer hence can only follow a straight swimming trajectory. Also notice that among all the modes in the squirming profile, Eq.~\ref{eqn:flowAxisBC}, just one mode  contributes to propulsion, namely mode $B_{01}$. Similarly, among all the new azimuthal modes in the boundary condition, only  mode $C_{01}$ contributes to the rotation of the squirmer. 

Finally, by  superimposing  the solution of the pumping problem, Eq.~\ref{eqn:PumpSwimg},  and the flow fields due to the swimming kinematics, Eqs.~\ref{eqn:AxisSpeed} and \ref{eqn:AxisRotation}, we  obtain the overall flow field of an axisymmetric swimming squirmer, $\v = v_r \e_r + v_\theta \e_\theta + v_\phi \e_\phi$,  in the laboratory frame as
\begin{eqnarray}
v_r   &=&   - \frac{4\cos \theta}{3r^3} B_{01} +\sum_{n=2}^\infty  \frac{(n+1) P_n}{  r^{n+2}} \left( \frac{r^2}{a^2}-1 \right) B_{0n}, \label{eqn:AxisSuperFlowR}\\
v_\theta &=& -\frac{2 \sin \theta}{3 r^3} B_{01} + \sum_{n=2}^\infty  \sin \theta P^{'}_n   \left( \frac{n-2}{n a^2 r^n}-\frac{1}{r^{n+2}} \right) B_{0n}, \\
v_\phi  &=&  \sum_{n=2}^\infty   \frac{\sin \theta P^{'}_n}{r^{n+1}}  C_{0n}  . \label{eqn:AxisSuperFlowPhi}
\end{eqnarray}
Note that throughout the paper we will refer to  the flow fields in the pumping and swimming problems as $\u$ and $\v$ respectively.

\subsection{The axisymmetric flow structure}\label{sec:AxisPhysics}

In this section, we identity the flow structure generated by a swimming squirmer as due to a superposition of flow singularities. This allows a physical interpretation of the flows caused by different modes of ciliary action in terms of combinations of point forces and torques and their spatial derivatives \cite{Ishikawa2006}. Such an understanding is useful for constructing approximations for swimmers in theoretical modeling and computer simulations, where one can retain only modes relevant to the aspects of physics of interest. For instance, in the axisymmetric case, it is common to retain only the mode contributing to swimming (the source dipole mode) and the mode due to two point forces (Stokes dipoles) \cite{Ishikawa2006,Ishikawa2007, Ishikawa2007b,Magar2003}, where the arrangement of the two point forces (the sign of the Stokes dipole) can represent different types of swimmers (``pushers" vs.~``pullers", see Sec.~\ref{sec:Intro}). We first revisit below the  known correspondences between the flow field around an axisymmetric squirmer and different flow singularities. We then proceed to discuss the new axisymmetric modes and the interpretation of their corresponding flow singularities.

\subsubsection{The $B_{01}$ mode} \label{sec:B01}

{The primary fundamental singularity in Stokes flows is  the flow due to a point force $f \alphaB \delta(\r)$ of magnitude $f$ and direction $\alphaB$ at the origin, where $\delta(\r)$ is the dirac delta function. The solution is given by $\u = f \G(\alphaB)/ (8\pi \eta)$, where
\begin{eqnarray}\label{eqn:Stokeslet}
\G(\alphaB) = \frac{1}{r} \left[ \alphaB + (\alphaB \cdot \e_r) \e_r \right],
\end{eqnarray}
and is called a Stokeslet.} That flow is long-ranged and decays as $1/r$. A Stokeslet acting in the $z$-direction has the explicit form in spherical coordinates
\begin{eqnarray}
\G(\e_z) = \frac{1}{r} \left[  2 \cos \theta \e_r - \sin \theta \e_\theta  \right] . \label{eqn:StokesletZ}
\end{eqnarray}

In the pumping problem, Eq.~\ref{eqn:flowAxis}, we can then identify that the $B_{01}$ mode 
\begin{eqnarray}
\u_{B_{01}} &=& \left(\frac{2P_1}{a^2r}- \frac{2P_1}{r^3}\right) B_{01} \e_r + \left(-\frac{\sin\theta P_1^{'}}{a^2r}- \frac{\sin\theta P_1^{'}}{r^3}    \right) B_{01} \e_\theta\\
&=& \frac{B_{01}}{a^2r} \left(2\cos\theta \e_r-\sin\theta \e_\theta \right) - \frac{B_{01}}{r^3} \left( 2\cos\theta \e_r+\sin\theta \e_\theta \right) ,
\end{eqnarray}
contains a Stokeslet in the $z$-direction (Eq.~\ref{eqn:StokesletZ}). The other component decaying faster as $1/r^3$ corresponds to a source dipole singularity, also in the $z$-direction, which we discuss below (see Eq.~\ref{eqn:PotentialDipoleZ}). 

To obtain the overall flow field, $\v$, surrounding a swimming squirmer, Eq.~\ref{eqn:PumpSwimg}, the  solution to the pumping problem, $\u$, needs to be superimposed with that due to a translating sphere at a velocity $-4B_{01}/3a^3$, Eq.~\ref{eqn:AxisSpeed}, leading to
\begin{eqnarray}
\u_T = -\frac{B_{01}}{a^2r} \left(2\cos\theta \e_r-\sin\theta \e_\theta \right) + \frac{B_{01}}{3r^3}\left(2\cos\theta\e_r+\sin\theta \e_\theta \right),
\end{eqnarray}
which also contains a Stokeslet and a source dipole. Unsurprisingly, the Stokeslet components cancel each other exactly and satisfy the overall force-free condition of a free swimming squirmer. Therefore, a Stokeslet component does not appear in the swimming flow field from Eqs.~\ref{eqn:AxisSuperFlowR}--\ref{eqn:AxisSuperFlowPhi}. Notice however that the cancellation of the source dipole components is incomplete, leaving a residual source dipole in the swimming flow field as
\begin{eqnarray}
\v_{B_{01}} = -\frac{2 B_{01}}{3r^3} \left(2\cos\theta \e_r + \sin\theta \e_\theta  \right) . \label{eqn:AxisResidual}
\end{eqnarray}

\subsubsection{The $B_{02}$ and $C_{01}$ modes}

Analyzing the structure of flow around a swimming squirmer from Eqs.~\ref{eqn:AxisSuperFlowR}--\ref{eqn:AxisSuperFlowPhi}, we see that the slowest decaying flow field ($\sim 1/r^2$) is contained in the $B_{02}$ mode, as
\begin{eqnarray}
\v_{B_{02}} =&& \frac{3 P_2}{r^{4}} \left( \frac{r^2}{a^2}-1\right) B_{02} \e_r - \frac{\sin\theta P_2^{'}}{r^4} B_{02} \e_\theta\\
=&& \frac{3 B_{02}}{4a^2 r^2}(1+3\cos2\theta) \e_r \nonumber \\
&-& \frac{3B_{02}}{4r^4} \left[ (1+3\cos2\theta)\e_r+2\sin2\theta \e_\theta \right] . \label{eqn:AxisB02}
\end{eqnarray}
The part decaying as $1/r^2$ can be be interpreted as the contribution of a Stokes dipole, which is a higher order singularity of Stokes flows and obtained by taking a derivative of a Stokeslet (directed in the $\alphaB$ direction) along the direction $\betaB$
\begin{eqnarray}\label{eqn:SD}
\G_D(\betaB, \alphaB) &=& \betaB \cdot \nabla \G(\alphaB) \nonumber \\
&=& \frac{(\betaB \times \alphaB) \e_r}{r^2}- \frac{(\betaB \cdot \alphaB) \e_r   - 3(\alphaB \cdot \e_r) (\betaB \cdot \e_r) \e_r}{r^2}\cdot
\end{eqnarray}
The symmetric part of a Stokes dipole is termed a stresslet, first defined by Batchelor \cite{Batchelor1970b}, and given by
\begin{eqnarray}
\SS(\betaB, \alphaB) = - \frac{(\betaB \cdot \alphaB) \e_r   - 3(\alphaB \cdot \e_r) (\betaB \cdot \e_r) \e_r}{r^2}, \label{eqn:DefStresslet}
\end{eqnarray}
which physically represents straining motion of the fluid. The antisymmetric part is termed a rotlet
\begin{eqnarray}\label{eqn:rotlet}
\R(\gammaB) = \frac{\zetaB \times \e_r}{r^2},
\end{eqnarray}
where $\zetaB = \betaB \times \alphaB$ represents the strength (magnitude and direction) of the flow due to a singular point torque. The Stokes dipole with $\alphaB = \betaB = \e_z$ corresponds to only a stresslet 
\begin{eqnarray}
\G_D(\e_z,\e_z) = \SS(\e_z,\e_z) = \frac{1+3\cos2\theta}{r^2} \e_r .
\end{eqnarray}
In the $B_{02}$ mode of the flow field, Eq.~\ref{eqn:AxisB02}, we can readily identify a Stokes dipole (stresslet)  {(see Appendices \ref{app:StokesDipole} \& \ref{app:AppendixStresslet})}, while the other part decaying as $ 1/r^4$ corresponds to a source quadrupole, a higher order singularity.

The first azimuthal mode in the pumping problem, Eq.~\ref{eqn:flowAxis}, is given by the $C_{01}$ mode
\begin{eqnarray}
\u_{C_{01}} = \frac{\sin \theta P_1^{'}}{r^2} C_{01} \e_\phi = \frac{\sin\theta}{r^2} C_{01} \e_\phi,
\end{eqnarray}
and represents a rotlet in the $z$-direction, $ \R(\e_z) = \sin\theta/r^2 \e_\phi$. Similar to the translation case, the pumping problem solution, $\u$, is superimposed with the flow field due to the induced rotation at the rate $-C_{01}/a^3$ (Eq.~\ref{eqn:AxisRotation})
\begin{eqnarray}
\u_R = - \frac{\sin\theta}{r^2} C_{01} .
\end{eqnarray}
leading to the total flow field surrounding a swimming squirmer, $\v$ (Eq.~\ref{eqn:PumpSwimg}). Again, unsurprisingly, the rotlet components exerting torques on the fluid cancel out completely thus satisfying the overall torque-free condition. The $C_{01}$ mode is hence absent from the resulting swimming flow field, Eq.~\ref{eqn:AxisSuperFlowR}--\ref{eqn:AxisSuperFlowPhi}, leaving no trace of the rotational motion of the squirmer. The important difference between rotation  and translation is that the rotational mode is due to velocities which are all in the direction of the rotation, so a complete cancellation of the flow field  satisfying the torque-free condition is possible by simply rotating in the opposite direction at the same rate. In contrast, for translational swimming, such exact cancellation is not possible because the surface velocity has a distribution of directions all along the sphere relative to its swimming direction. In other words, one can construct the ultimate stealth rotating sphere using purely tangential modes but not a similarly stealth translating sphere.

In summary, the $B_{02}$ and $C_{01}$ modes together contain the representation of a Stokes dipole (stresslet plus rotlet) with the direction and gradient taken both in the $z$-direction ($\alphaB = \betaB = \e_z$).

\subsubsection{The $B_{03}$ and $C_{02}$ modes}
Higher order flow singularities can be obtained by repeatedly taking derivatives of the lower order singularities. For example, a Stokes quadrupole can be obtained by taking a derivative of a Stokes dipole, $\G_D(\betaB, \alphaB)$, along the direction $\gammaB$, leading to
\begin{eqnarray}\label{eqn:GSQ}
\G_{Q}(\gammaB, \betaB, \alphaB) =&& \ \gammaB \cdot \nabla \G(\betaB,\alphaB) \nonumber \\
=&& \ \frac{1}{r^3} \Big\{ (\betaB \cdot \alphaB) \gammaB + (\gammaB \cdot \alphaB) \betaB-(\gammaB \cdot \betaB) \alphaB \nonumber \\
&+&15 (\gammaB \cdot \e_r) (\betaB \cdot \e_r) (\alphaB \cdot \e_r)\e_r  \nonumber\\
&-& 3 \big[ (\betaB \cdot \alphaB) (\gammaB \cdot \e_r) + (\gammaB \cdot \alphaB) (\betaB \cdot \e_r) + (\gammaB \cdot \betaB) (\alphaB \cdot \e_r) \big] \e_r \nonumber \\
&-&3 \big[ (\gammaB \cdot \e_r)(\alphaB \cdot \e_r) \gammaB+(\gammaB \cdot \e_r) (\alphaB \cdot \e_r)\betaB \nonumber \\
&-&(\gammaB \cdot \e_r) (\betaB \cdot \e_r) \alphaB \big] \Big\}.
\end{eqnarray}
The flow field due to such a Stokes quadrupole decays as $ 1/r^3$.  In particular, a Stokes quadrupole with $\alphaB = \betaB = \gammaB = \e_z$ takes the simple form
\begin{eqnarray}
\G_Q(\e_z,\e_z,\e_z) = \frac{1}{r^3} \left[ (\cos\theta+3\cos3\theta) \e_r +\frac{1}{4}(3\sin3\theta-\sin\theta) \e_\theta \right], \label{eqn:GQzzz}
\end{eqnarray}
which is useful in interpreting the $B_{03}$ mode in Lamb's solution as we will see below.

Several components of the Stokes quadrupole have particularly clear physical meanings, such as the potential (source) dipole
\begin{eqnarray} \label{eqn:podi}
\P_D(\alphaB) = \frac{1}{r^3} \left[ -\alphaB + 3(\alphaB \cdot \e_r) \e_r \right],
\end{eqnarray}
where $\alphaB$ denotes its direction. A potential dipole in the $z$-direction is given by
\begin{eqnarray}
\P_D(\e_z) = \frac{1}{r^3} \left[ 2 \cos \theta \ \e_r + \sin \theta \ \e_\theta \right],\label{eqn:PotentialDipoleZ}
\end{eqnarray}
which is the residual component of the $B_{01}$ mode in the overall flow field of a swimming squirmer (see Eq.~\ref{eqn:AxisResidual} or Eqs.~\ref{eqn:AxisSuperFlowR}--\ref{eqn:AxisSuperFlowPhi}). This potential dipole component contained in the $B_{01}$ mode, together with the part decaying as $ 1/r^3$ in the $B_{03}$ mode and given by
\begin{eqnarray}
\v_{B_{03}} =&& \ \frac{B_{03}}{2a^2r^3} \left[(3\cos\theta+5\cos3\theta) \e_r + \frac{1}{4} (\sin\theta+5\sin3\theta)\e_\theta \right] \nonumber \\
&-& \frac{B_{03}}{2r^5} \left[(3\cos\theta+5\cos3\theta) \e_r+ \frac{3}{4} (\sin\theta+5\sin3\theta)\e_\theta \right],
\end{eqnarray}
contain the representation of the Stokes quadrupole, $\G_Q(\e_z,\e_z,\e_z)$, expressed in Eq.~\ref{eqn:GQzzz}. The other component decaying as $1/r^5$ in the $B_{03}$  mode corresponds to a source octupole.

\begin{figure}
\centering
\includegraphics[width=1\textwidth]{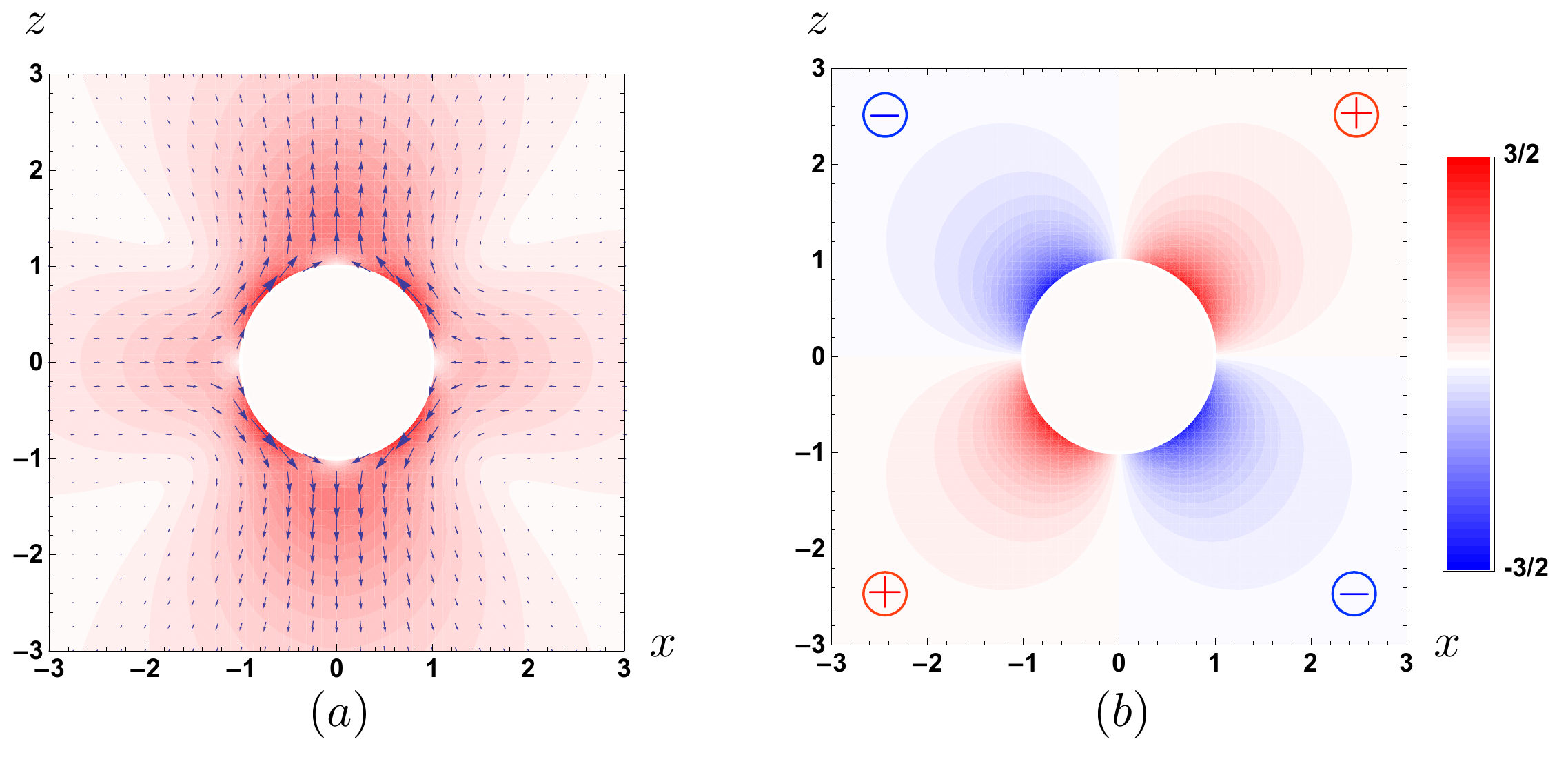}
\caption{Flow fields due to the axisymmetric squirming motion of a sphere of radius $a=1$. (a):  Flow field due to the $B_{02}$ mode, $\v_{B_{02}} = 3(1+3\cos2\theta)B_{02}/4a^2r^2 \e_r -3[(1+3\cos2\theta)\e_r + 2\sin2\theta \e_\theta] B_{02}/4r^4$, which is a source quadrupole in the near field, and a stresslet in the far field. The color density represents the flow speed. (b): Flow field due to the $C_{02}$ mode, $\v_{C_{02}} = 3\sin2\theta/2r^3 \e_\phi$ (a rotlet dipole), with the velocities all in the azimuthal ($\phi$) direction, i.e.~perpendicular to the paper). The color density represents the speed, with red and blue denoting positive ``+" (into page) and negative ``$-$" (out of page) velocities in different quadrants.}
\label{fdrd}
\end{figure}

The first azimuthal component appearing in the flow field around a swimming squirmer is given by the $C_{02}$ mode
\begin{eqnarray}
\v_{C_{02}} =\frac{3\sin2\theta }{2r^3} C_{02} \e_\phi, \label{eqn:C02}
\end{eqnarray}
which represents another well-known component of the Stokes quadrupole, named a rotlet dipole. A rotlet dipole can be obtained by taking a derivative along the $\gammaB$ direction of a rotlet with the direction $\zetaB = \betaB \times \alphaB$ 
\begin{eqnarray} \label{eqb:RotDi}
\R_D(\gammaB, \zetaB) = \gammaB \cdot \nabla \R(\zetaB) = \frac{1}{r^3} \left[ \gammaB \times \zetaB + \frac{3(\gammaB \cdot \e_r)(\zetaB \times \e_r)}{r^3} \right] \cdot
\end{eqnarray}
In particular, a rotlet dipole with $\gammaB=\zetaB=\e_z$ takes the simple form $\R_D(\e_z,\e_z) = 3 \sin 2\theta/2r^3\e_\phi$. This corresponds to the $C_{02}$ mode in Lamb's general solution, Eq.~\ref{eqn:C02}, and provides the leading-order mode in the azimuthal direction. In Fig.~\ref{fdrd}, we plot the slowest decaying flow field, given by the $B_{02}$ mode (a stresslet in the far field, Fig.~\ref{fdrd}a), as well as  the slowest decaying flow in the azimuthal direction, given by the $C_{02}$ mode (a rotlet dipole, Fig.~\ref{fdrd}b).

To summarize, the $B_{01}$ and $B_{03}$ modes contain physically the Stokes quadrupole, $\G_Q(\e_z,\e_z,\e_z)$; the $C_{02}$ mode corresponds to a rotlet dipole, $\R_D(\e_z,\e_z) = 3 \sin 2\theta/2r^3\e_\phi$, which is part of a Stokes quadrupole different than $\G_Q(\e_z,\e_z,\e_z)$ (see also Sec.~\ref{sec:NonAxisPhysics} for further details).

\section{Non-axisymmetric Squirming Motion}\label{sec:NonAxisMain}
We now generalize the results for axisymmetric swimming to the non-axisymmetric case  using Eqs.~\ref{eqn:urTan}--\ref{eqn:uphiTan}. The  velocities on the surface of the  sphere for this general case are given by
\begin{eqnarray}
u_r|_{r=a} =&& 0, \label{eqn:urTanNonA} \\
u_\theta|_{r=a}  =&& \sum_{n=1}^\infty \sum_{m=0}^n  \bigg [   -\frac{2\sin\theta P^{m'}_n}{n a^{n+2}}   \left( B_{mn} \cos m\phi +  \tilde{B}_{mn} \sin m\phi \right) \nonumber \\
&+& \frac{mP^m_n}{a^{n+1} \sin \theta}  (\tilde{C}_{mn} \cos m\phi - C_{mn} \sin m \phi )  \bigg ] \label{eqn:uthetaTanNonA},\\
u_\phi|_{r=a}  =&& \sum_{n=1}^\infty \sum_{m=0}^n \bigg[ \frac{\sin \theta P^{m'}_n}{a^{n+1}}  (C_{mn} \cos m \phi + \tilde{C}_{mn} \sin m \phi)  \nonumber \\
&+& \frac{2mP^m_n}{na^{n+2}\sin \theta} \left(\tilde{B}_{mn} \cos m \phi - B_{mn} \sin m \phi \right) \bigg ]  \label{eqn:uphiTanNonA}.
\end{eqnarray}
Recall that the more general  analysis which includes nonzero radial surface velocities is given in Appendices \ref{app:AppendixSwimming}--\ref{app:AppendixWork} and we focus below on the swimming problem of a non-axisymmetric squirmer with purely tangential squirming profiles.

\subsection{Swimming of a non-axisymmetric squirmer}\label{sec:NonAxis}

We follow closely the analysis presented in the axisymmetric case (Sec.~\ref{sec:Axis}) to investigate the situation where the surface motion is non-axisymmetric. As discussed in Sec.~\ref{sec:AxisSwimming}, we consider the swimming problem as a superposition of a pumping problem with the boundary actuation in Eqs.~\ref{eqn:urTanNonA}--\ref{eqn:uphiTanNonA} and the flow field due to the induced translation and rotation of the squirmer (given in Eq.~\ref{eqn:PumpSwimg}). Applying the force and torque balances in this non-axisymmetric case, the results from  Eqs.~\ref{eqn:forceBalanceD} and \ref{eqn:torqueBalanceD} continue to hold, but with  solutions for $p_{-2}$ and $\chi_{-2}$ which are more involved, resulting in
{\begin{eqnarray}
\U &=&- \frac{2}{3 \eta a} \nabla\left[ r \left( P_1 A_{01}+ P_1^1 \cos \phi A_{11}+ P_1^1 \sin \phi \tilde{A}_{11} \right)  \right] \nonumber \\
&=& \frac{4}{3a^3} \left( B_{11} \e_x + \tilde{B}_{11} \e_y - B_{01} \e_z \right), \label{eqn:NonAxisSpeed}\\
\OmegaB &=&- \frac{1}{a^3} \nabla \left[ r \left( P_1 C_{01}+ P_1^1 \cos \phi C_{11}+ P_1^1 \sin \phi \tilde{C}_{11} \right)  \right]  \nonumber \\
&=& \frac{1}{a^3} \left(C_{11}\e_x + \tilde{C}_{11} \e_y - C_{01} \e_z \right) . \label{eqn:NonAxisRotSpeed}
\end{eqnarray}}

For an axisymmetric squirmer, only the $B_{01}$ and $C_{01}$ modes contribute to propulsion and rotation respectively. In contrast, for a general non-axisymmetric squirmer, Eqs.~\ref{eqn:NonAxisSpeed}-\ref{eqn:NonAxisRotSpeed} identify all the modes contributing to its three-dimensional locomotion. Specifically, three modes, $B_{11}$, $\tilde{B}_{11}$, and $B_{01}$, contribute to the translational swimming in the $x$, $y$, and $z$-directions respectively; similarly, three modes, $C_{11}$, $\tilde{C}_{11}$, and $C_{01}$, lead to rotation in the $x$, $y$, and $z$-directions respectively. By  superimposing the flow fields due to the induced translation and rotation according to the velocities determined, Eqs.~\ref{eqn:NonAxisSpeed}--\ref{eqn:NonAxisRotSpeed}, with the solution to the non-axisymmetric pumping problem, Eqs.~\ref{eqn:urTan}--\ref{eqn:uphiTan}, we  obtain the flow field around a non-axisymmetric swimming squirmer  as
\begin{eqnarray}
v_r =&& \ \frac{4}{3 r^3} \left( B_{11} \sin \theta \cos \phi + \tilde{B}_{11} \sin \theta \sin \phi - B_{01} \cos \theta   \right) \nonumber \\
&+&\sum_{n=2}^\infty \sum_{m=0}^n \frac{(n+1) P^m_n}{r^{n+2}} \left( \frac{r^2}{a^2}- 1 \right) \left[  B_{mn}\cos m \phi +  \tilde{B}_{mn} \sin m \phi   \right], \label{eqn:NonAxisSwimFlowR}\\
v_\theta =&& - \frac{2}{3 r^3} \left(  B_{11} \cos \theta \cos \phi + \tilde{B}_{11} \cos \theta \sin \phi +B_{01} \sin \theta  \right)  \nonumber \\
&+& \sum_{n=2}^\infty \sum_{m=0}^n  \bigg [   \sin \theta P^{m'}_n   \left( \frac{n-2}{n a^2 r^n}-\frac{1}{r^{n+2}} \right) \left( B_{mn} \cos m\phi +  \tilde{B}_{mn} \sin m\phi \right)  \nonumber \\
&+&  \frac{mP^m_n}{r^{n+1}\sin \theta} (\tilde{C}_{mn} \cos m\phi - C_{mn} \sin m \phi )  \bigg ], \\
v_\phi =&& \ \frac{2}{3 r^3} \left( B_{11} \sin \phi-\tilde{B}_{11} \cos \phi \right) \nonumber \\
&+& \sum_{n=2}^\infty \sum_{m=0}^n \bigg[ \frac{\sin \theta P^{m'}_n }{r^{n+1}} (C_{mn} \cos m \phi + \tilde{C}_{mn} \sin m \phi)  \nonumber \\
&-&  \frac{mP^m_n}{\sin \theta} \left( \frac{n-2}{na^2 r^n}-\frac{1}{r^{n+2}}\right) \left(\tilde{B}_{mn} \cos m \phi - B_{mn} \sin m \phi \right) \bigg ] \cdot \label{eqn:NonAxisSwimFlowPhi}
\end{eqnarray}
The flow reduces to the Eqs.~\ref{eqn:AxisSuperFlowR}--\ref{eqn:AxisSuperFlowPhi} in the axisymmetric case ($m=0$). 
The physical meaning of the new non-axisymmetric terms ($m\neq0$) is interpreted in Sec.~\ref{sec:NonAxisPhysics}.

\subsection{Swimming kinematics by integral theorems}\label{sec:reciprocal}

The swimming kinematics of a squirmer can also be arrived using the reciprocal theorem approach taken by Stone \& Samuel \cite{Stone1996} without having to solve for the whole flow field. The theorem relates the swimming velocity to the surface distortion, $\u|_{r=a}$, via a surface integral on the sphere S
\begin{eqnarray}
\U = - \frac{1}{4 \pi a^2} \int_{\textrm{S}} \u|_{r=a} dS,
\end{eqnarray}
which in spherical coordinates reads
\begin{eqnarray}
\U  = -\frac{1}{4 \pi} \int_0^{2\pi} \int_{-1}^1 (u_\theta \e_\theta   +u_\phi \e_\phi)_{r=a}    d\mu d\phi  .
\end{eqnarray}
By transforming the basis vectors in spherical coordinates to those in Cartesian coordinates, the integral simplifies due to the orthogonality of sinusoidal functions in the azimuthal angle $\phi$, and we obtain
{\begin{eqnarray}
\U  =&& -\frac{1}{4}  \sum_{n=1}^\infty \Bigg\{ \frac{\tilde{C}_{1n} }{a^{n+1}} \int_{-1}^1 \left(P_1^{1'}P_n^1+P_1^1 P_n^{1'} \right)d\mu \nonumber \\
&-& \frac{2B_{1n}}{na^{n+2}} \int_{-1}^1 \left[(1-\mu^2) P_1^{1'} P_n^{1'}+\frac{P_1^1 P_n^1}{1-\mu^2} \right] d\mu d\phi   \Bigg\} \e_x \nonumber\\
&-&\frac{1}{4}\sum_{n=1}^\infty \Bigg\{ \frac{C_{1n}}{a^{n+1}} \int_{-1}^1 \left(P_1^{1'}P_n^1+P_1^1 P_n^{1'} \right) d\mu \nonumber \\
&-& \frac{2\tilde{B}_{1n} }{na^{n+2}} \int_{-1}^1 \left[(1-\mu^2) P_1^{1'} P_n^{1'}+\frac{P_1^1 P_n^1}{1-\mu^2} \right]  d\mu d\phi \Bigg\} \e_y \nonumber \\
&-&\frac{1}{2} \sum_{n=1}^\infty \left[ \frac{2 B_{0n} }{a^{n+2}n} \int_{-1}^1(1-\mu^2) P_1^{'} P_n^{'} d\mu\right] \e_z .
\end{eqnarray}}
The integrals associated with $C_{1n}$ and $\tilde{C}_{1n}$ vanish upon integration by parts. The remaining integrals can be evaluated using a general expression  derived below (see Eq.~\ref{eqn:Useful})  in the special cases of $m=0$ and 1. We then obtain the results identical to those given by Lamb's solution, Eq.~\ref{eqn:NonAxisSpeed}. Similarly, one can employ the reciprocal theorem to derive the rotational rotational velocity \cite{Stone1996}
\begin{eqnarray}
\OmegaB  = - \frac{3}{8 \pi a^3} \int_{\textrm{S}} \n \times \u|_{r=a} dS,
\end{eqnarray}
and obtain the same result as above (Eq.~\ref{eqn:NonAxisRotSpeed}). However, while the reciprocal theorem is a useful tool for determining  swimming kinematics, it provides no information about the flow around the swimmer, which is a main result of our work.

\subsection{The non-axisymmetric flow structure}\label{sec:NonAxisPhysics}
In the axisymmetric case, we have interpreted the flow fields due to different modes of the squirming profile as fundamental flow singularities. We extend the idea here to physically interpret the flow induced by the non-axisymmetric terms. Note that, in each of the modes discussed below, the corresponding flow field is not a far-field approximation of the  flow induced by the squirmer but an exact solution valid in the entire space, and it is an appropriate superposition of these modes which satisfies arbitrary boundary conditions on the spherical surface.

\subsubsection{The $B_{11}$ and $\tilde{B}_{11}$ modes}
In  Sec.~\ref{sec:AxisPhysics}, we have identified that part of the $B_{01}$ mode of Lamb's solution in the pumping problem, Eqs.~\ref{eqn:urTanNonA}--\ref{eqn:uphiTanNonA}, corresponds to a Stokeslet directed in the $z$-direction. One can then verify that other non-axisymmetric modes decaying as $1/r$ in Lamb's solution, namely $B_{11}$ and $\tilde{B}_{11}$, with the flow fields
\begin{eqnarray}
\u_{B_{11}} =&& \ \frac{B_{11}}{a^2r} \left( -2\cos\theta \cos\phi \ \e_r -\cos\theta\cos\phi \ \e_\theta + \sin\phi \ \e_\phi \right)  \nonumber \\
&+& \frac{B_{11}}{r^3} \left( 2\sin\theta\cos\phi \ \e_r -\cos\theta\cos\phi \ \e_\theta + \sin\phi \ \e_\phi  \right), \\
\u_{\tilde{B}_{11}} =&& \ \frac{\tilde{B}_{11}}{a^2r} \left( -2\sin\theta\sin\phi \ \e_r -\cos\theta \sin\phi \ \e_\theta - \cos\phi \ \e_\phi \right) \nonumber \\
&+& \frac{\tilde{B}_{11}}{r^3} \left( 2\sin\theta\sin\phi \ \e_r - \cos\theta\sin\phi \ \e_\theta -\cos\phi \ \e_\phi \right),
\end{eqnarray}
contain Stokeslets directed in the $x$ and $y$-directions respectively (see Appendix \ref{app:Stokeslet}). Similarly, the parts decaying as $1/r^3$ in the $B_{11}$ and $\tilde{B}_{11}$ modes correspond to potential dipoles in the $x$ and $y$-directions respectively (see Appendix \ref{app:PotentialDipole}).

Naturally, all  Stokeslet components are cancelled out exactly upon the superposition with the flow fields due to translational swimming in different directions, Eq.~\ref{eqn:NonAxisSpeed}, satisfying the force-free condition and leaving only residual potential dipoles in different directions in the flow field of a non-axisymmetric swimming squirmer, Eqs.~\ref{eqn:NonAxisSwimFlowR}--\ref{eqn:NonAxisSwimFlowPhi}. 

\subsubsection{The $B_{m2}$, $\tilde{B}_{m2}$, $C_{11}$, and $\tilde{C}_{11}$ modes ($1\le m \le2$)}

Once the Stokeslet contributions have all been removed, the slowest decaying components in the flow around a general (non-axisymmetric) swimming squirmer, Eqs.~\ref{eqn:NonAxisSwimFlowR}--\ref{eqn:NonAxisSwimFlowPhi},  decay as $1/r^2$, which are associated with the $B_{02}$, $B_{m2}$, $\tilde{B}_{m2}$ modes ($1\le m \le2$). The parts decaying as $1/r^2$ in the non-axisymmetric terms ($B_{m2}$ and $\tilde{B}_{m2}$) have the same physical meaning as that in the axisymmetric $B_{02}$ mode -- these are stresslets -- but they are formed by taking a gradient along various directions of a Stokeslet, itself aligned in different directions. Specifically, the $B_{12}$ mode alone contains the stresslet $\SS(\e_z,\e_x)$. The $\sim 1/r^2$ components in the $B_{22}$ and $B_{02}$ modes can be combined to represent the stresslet $\SS(\e_x, \e_x)$, formed by taking the gradient along $\e_x$ of a Stokeslet directed also in $\e_x$. Alternatively, one can understand the component decaying as $1/r^2$ in the $B_{22}$ mode alone represents a stresslet formed by the superposition of $\SS(\e_x, \e_x)$ and $-\SS(\e_z, \e_z)$. We refer to Appendix \ref{app:AppendixStresslet} for the expressions of all stresslets in different configurations. The correspondence between stresslets with different configurations and the modes in Lamb's general solution is summarized in Table \ref{table:ForceDipoles}.

\begin{table}[t]
\small
  \caption{Correspondence between  force dipoles (stresslets plus rotlets) and the different modes of the squirming motion.}
  \label{table:ForceDipoles}
  \centering
  \begin{tabular*}{1\textwidth}{@{\extracolsep{\fill}}ccp{1.6cm}ccc}
   \hline
   Force Dipoles  & Contained & Stresslets  & Contained & Rotlets  & Contained \\
   ($\sim 1/r^2$) & in modes & ($\sim 1/r^2$)  & in modes & ($\sim 1/r^2$)  & in modes \\
    \hline
    $\G_D(\e_x, \e_x)$ & $B_{02}, B_{22}$ & $\SS(\e_x,\e_x)=\G_D(\e_x, \e_x)$ & $B_{02}$, $B_{22}$ &$\R_D(\e_x)$& $C_{11}$\\
    $\G_D(\e_y, \e_x)$ & $\tilde{B}_{22}, C_{01}$& $\SS(\e_y,\e_x) = \SS(\e_x,\e_y)$ & $\tilde{B}_{22}$& $\R_D(\e_y)$ & $ \tilde{C}_{11}$  \\
    $\G_D(\e_z, \e_x)$ & $B_{12}, \tilde{C}_{11}$&$\SS(\e_z,\e_x) = \SS(\e_x,\e_z)$  & $B_{12}$ &$\R_D(\e_z)$ & $C_{01}$\\
    $\G_D(\e_x, \e_y)$ & $\tilde{B}_{22}, C_{01}$&$\SS(\e_y,\e_y)=\G_D(\e_y,\e_y)$  &$B_{02}, B_{22}$ & & \\
    $\G_D(\e_y, \e_y)$ & $B_{02}, B_{22}$& $\SS(\e_z,\e_y) = \SS(\e_z,\e_y)$& $\tilde{B}_{12}$&& \\
    $\G_D(\e_z, \e_y)$ & $\tilde{B}_{12}, C_{11}$&$\SS(\e_z,\e_z)=\G_D(\e_z,\e_z)$ &$B_{02}$ && \\
    $\G_D(\e_x, \e_z)$ & $B_{12}, \tilde{C}_{11}$ & &&&\\
    $\G_D(\e_y, \e_z)$ & $\tilde{B}_{12}, C_{11}$& &&&\\
    $\G_D(\e_z, \e_z)$ & $B_{02}$ & & &&\\
    \hline
  \end{tabular*}
\end{table}

As first proposed by Batchelor  \cite{Batchelor1970b}, one can  also write down a stresslet tensor in the Cartesian coordinates containing the contributions from different modes of squirming motion, and we obtain
\begin{eqnarray}
\mathbf{S} = - \frac{8 \pi \eta}{a^2} 
\left( \begin{array}{ccc}
-\frac{B_{02}}{2} + 3 B_{22} & 3 \tilde{B}_{22} & -\frac{3}{2} B_{12} \\
3 \tilde{B}_{22} &  -\frac{B_{02}}{2}-3B_{22} & -\frac{3}{2} \tilde{B}_{12} \\
-\frac{3}{2} B_{12} & -\frac{3}{2} \tilde{B}_{12} & B_{02}\end{array} \right) \cdot
\end{eqnarray}
That equation, allowing to determine the exact far-field nature of a generalized squirmer, is one of the important results of our paper. 

In the pumping solution, Eqs.~\ref{eqn:urTan}--\ref{eqn:uphiTan}, the $C_{02}$, $C_{11}$, and $\tilde{C}_{11}$ mode in Lamb's solution also decay as $1/r^2$. They however do not contribute to the overall flow field in the swimming problem due to the torque-free condition. Similar to the $C_{02}$ mode, the $C_{11}$ and $\tilde{C}_{11}$ modes represent rotlets in the $x$ and $y$-directions respectively. The expressions of rotlets with different configurations are reproduced in  Appendix \ref{app:AppendixRotlet}.

\subsubsection{The $B_{q3}$, $\tilde{B}_{q3}$, $C_{m2}$, and $\tilde{C}_{m2}$ modes ($1\le q \le 3$ and $1\le m \le 2$)}

In a similar fashion, one can identify that the $B_{q3}$, $\tilde{B}_{q3}$, $C_{m2}$, and $\tilde{C}_{m2}$ modes ($1\le q \le 3$ and $1\le m \le 2$) have the same physical meaning as their counterparts in the axisymmetric case (namely the $B_{03}$ and $C_{02}$ modes). Jointly with the parts decaying as $1/r^3$ in the $B_{01}$, $B_{11}$, and $\tilde{B}_{11}$ modes, the six modes $B_{03}$, $B_{q3}$, $\tilde{B}_{q3}$, $C_{02}$, $C_{m2}$, and $\tilde{C}_{m2}$   contain the representation of a general Stokes quadrupole with all possible geometrical configurations. Which mode corresponds to which quadrupole is  summarized in Table \ref{table:ForceQuad}.

\begin{table}[h]
\small
  \caption{Correspondence between the force quadrupoles and the different modes of the squirming motion.}
  \label{table:ForceQuad}
  \centering
  \begin{tabular*}{1\textwidth}{@{\extracolsep{\fill}}lc}
    \hline
   Force Quadrupoles ($\sim 1/r^3$) & Contained in modes \\
    \hline
    $\G_Q(\e_x, \e_x, \e_x)$ & $B_{11}, B_{13}, B_{33}$\\
        $\G_Q(\e_y, \e_x, \e_x)=\G_Q(\e_x, \e_y, \e_x)$ & $\tilde{B}_{11}, \tilde{B}_{13}, \tilde{B}_{33}, C_{12}$ \\
            $\G_Q(\e_z, \e_x, \e_x)=\G_Q(\e_x, \e_z, \e_x)$ & $B_{01}, B_{03}, B_{23}, \tilde{C}_{22}$ \\
                    $\G_Q(\e_y, \e_y, \e_x)$ & $B_{11}, B_{13}, B_{33}, \tilde{C}_{12}$ \\
                        $\G_Q(\e_z, \e_y, \e_x)=\G_Q(\e_y, \e_z, \e_x)$ & $\tilde{B}_{23}, C_{22}, C_{02}$ \\
                            $\G_Q(\e_z, \e_z, \e_x)$ & $B_{11}, B_{13}, \tilde{C}_{12}$ \\
    $\G_Q(\e_x, \e_x,\e_y)$ & $\tilde{B}_{11}, \tilde{B}_{13}, \tilde{B}_{33}, C_{12}$  \\
        $\G_Q(\e_y, \e_x, \e_y)=\G_Q(\e_x, \e_y, \e_y)$ & $B_{11}, B_{13}, B_{33}, \tilde{C}_{12}$ \\
            $\G_Q(\e_z, \e_x, \e_y)=\G_Q(\e_x, \e_z, \e_y)$ & $\tilde{B}_{23}, C_{22}, C_{02}$\\
                $\G_Q(\e_y, \e_y, \e_y)$ & $\tilde{B}_{11}, \tilde{B}_{13}, \tilde{B}_{33}$ \\
                    $\G_Q(\e_z, \e_y, \e_y)=\G_Q(\e_y, \e_z, \e_y)$ & $B_{01},B_{03},B_{23}, \tilde{C}_{22}$ \\
                        $\G_Q(\e_z, \e_z, \e_y)$ & $\tilde{B}_{11}, \tilde{B}_{13}, C_{12}$ \\
    $\G_Q(\e_x, \e_x, \e_z)$ & $B_{01}, B_{03}, B_{23}, \tilde{C}_{22}$ \\
    $\G_Q(\e_y, \e_x, \e_z)=\G_Q(\e_x, \e_y, \e_z)$ & $\tilde{B}_{23}, C_{22}$ \\
    $\G_Q(\e_z, \e_x, \e_z)=\G_Q(\e_x, \e_z, \e_z)$ & $B_{11}, B_{13}, \tilde{C}_{12}$ \\
    $\G_Q(\e_y, \e_y, \e_z)$ & $B_{01}, B_{03}, B_{23}, \tilde{C}_{22}$ \\
    $\G_Q(\e_z, \e_y, \e_z)=\G_Q(\e_y, \e_z, \e_z)$ & $\tilde{B}_{11}, \tilde{B}_{13}, C_{12}$ \\
    $\G_Q(\e_z, \e_z, \e_z)$ & $B_{01}, B_{03}$ \\
    \hline
  \end{tabular*}
\end{table}

In particular, we can identify the modes associated with better known components of the Stokes quadrupole, namely the potential dipoles and the rotlet dipoles. The flow fields associated with modes $B_{11}$ and $\tilde{B}_{11}$, and decaying as $1/r^3$, correspond to potential dipoles in the $x$ and $y$-directions respectively (see Table \ref{table:RotletDipoles} and Appendix \ref{app:PotentialDipole}). The $C_{02}$, $C_{m2}$, and $\tilde{C}_{m2}$ modes with the potential dipole modes ($B_{01}$, $B_{11}$ and $\tilde{B}_{11}$) contain the representation of a general three-dimensional rotlet dipole with different configurations (see Table \ref{table:RotletDipoles} and Appendix \ref{app:RotletDipole}).

\begin{table}
\small
  \caption{Correspondence between both rotlet dipoles and potential dipoles and the different modes of the squirming motion.}
  \label{table:RotletDipoles}
  \centering
  \begin{tabular*}{1\textwidth}{@{\extracolsep{\fill}}cp{1.5cm}cp{1.5cm}}
    \hline
   Rotlet Dipoles ($\sim 1/r^3$) & Contained in modes & Source Dipole ($\sim 1/r^3$) & Contained in modes\\
    \hline
    $\R_D(\e_x, \e_x)$ & $C_{02}, C_{22}$ & $\P_D (\e_x)$ & $B_{11}$ \\
      $\R_D(\e_x, \e_y)$ & $B_{01}, \tilde{C}_{22}$  & $\P_D(\e_y)$ & $\tilde{B}_{11}$ \\
    $\R_D(\e_x, \e_z)$ & $\tilde{B}_{11}, C_{12}$ & $\P_D(\e_z)$ & $B_{01}$\\
    $\R_D(\e_y, \e_x)$ & $B_{01}, \tilde{C}_{22}$ & & \\
    $\R_D(\e_y, \e_y)$ & $C_{02}, C_{22}$& & \\
    $\R_D(\e_y, \e_z)$ & $B_{11}, \tilde{C}_{12}$ & & \\
    $\R_D(\e_z, \e_x)$ & $\tilde{B}_{11}, C_{12}$ & &\\
    $\R_D(\e_z, \e_y)$ & $B_{11}, \tilde{C}_{12}$&  &\\
    $\R_D(\e_z, \e_z)$ & $C_{02}$ & &\\
    \hline
  \end{tabular*}
\end{table}

At this point it should thus be clear   that the flow field generated by a swimming squirmer, Eqs.~\ref{eqn:NonAxisSwimFlowR}--\ref{eqn:NonAxisSwimFlowPhi}, may be also viewed as combinations of fundamental flow singularities. The non-axisymmetric terms have the same physical meanings as their counterparts in the axisymmetric case (with the same value of $n$ in Eqs.~\ref{eqn:urTan}--\ref{eqn:uphiTan}), but they  include all  possible configurations of the flow singularities (for different values of $m$ in Eqs.~\ref{eqn:urTan}--\ref{eqn:uphiTan}). From a physical point of view, the essential physics is therefore all contained in the case of axisymmetric squirming motion and the non-axisymmetric flow fields are linear superpositions of flow singularities in different directions. For a given non-axisymmetric squirming profile, the more general analysis in our  paper provides a way to quantify, and understand, the general three-dimensional flow structure.

Finally, we note   that in the axisymmetric case, Sec.~\ref{sec:NonAxis}, the translational and rotational velocities are always in the same direction, and hence an axisymmetric squirmer can only swim along a straight line (possibly in an unsteady fashion). This is generalized in the non-axisymmetric case, where the motion of a steady squirmer is in general helical provided that $\U \cdot \OmegaB \neq 0$. The special case 
\begin{eqnarray}
\U \cdot \OmegaB = 0, \label{eqn:circle}
\end{eqnarray}
reduces the helical trajectory to a circle (a helix with zero pitch, Eq.~\ref{eqn:circle}) 
while in the situation where 
\begin{equation}
\U \times \OmegaB = \mathbf{0}, \label{eqn:line}
\end{equation}
the trajectory is reduced to  a straight line (a helix with zero radius, Eq.~\ref{eqn:line}). According to the swimming kinematics computed above in  Eqs.~\ref{eqn:NonAxisSpeed}--\ref{eqn:NonAxisRotSpeed}, and using Eqs.~\ref{eqn:uthetaTanNonA}--\ref{eqn:uphiTanNonA}, we obtain that a squirmer performs circular motion when
\begin{eqnarray}
B_{11} C_{11} + \tilde{B}_{11} \tilde{C}_{11} + B_{01} C_{01} &= 0,
\end{eqnarray}
while it follows a straight line when
\begin{eqnarray}
\tilde{B}_{11} C_{01} - B_{01} \tilde{C}_{11} = B_{11} C_{01} - B_{01}C_{11} &= B_{11} \tilde{C}_{11} - C_{11} \tilde{B}_{11} = 0 .
\end{eqnarray}

\subsection{Rate of Work}\label{sec:Rate}
In this section, the rate of working by the surface, $\mathcal{P}$, during the squirming motion is considered. In spherical coordinates, we write
\begin{eqnarray}
\mathcal{P} &=& -\int_S \n \cdot \sigmaB \cdot \v dS \nonumber \\
&=& - \int_0^{2\pi} \int_{0}^\pi  \left(\sigma_{rr} v_r + \sigma_{r \theta} v_\theta + \sigma_{r\phi} v_\phi\right)_{r=a} a^2 \sin \theta d\theta d\phi. \label{eqn:P}
\end{eqnarray} 
The integrand can be evaluated with the Newtonian constitutive relation, $\sigmaB = -p + \eta(\nabla \v^T + \nabla \v)$, and the overall flow field of a swimming squirmer  from Eqs.~\ref{eqn:NonAxisSwimFlowR}--\ref{eqn:NonAxisSwimFlowPhi}. After some lengthy manipulation, we find that, for purely tangential deformation, the rate of work for general non-axisymmetric squirming motion is given by the positive-definite formula
\begin{eqnarray}
\mathcal{P} =&& \frac{64\pi \eta}{3a^5}\left( B_{01}^2+B_{11}^2+\tilde{B}_{11}^2 \right) \nonumber \\
&+&\sum_{n=2}^\infty   \frac{4n(n+1) \pi \eta}{a^{2n+1}} \left(  \frac{4}{n^2 a^2} B_{0n}^2 + \frac{n+2}{2n+1} C_{0n}^2 \right) \nonumber \\
&+& \sum_{n=2}^\infty \sum_{m=1}^n  \frac{ 2n(n+1)(n+m)! \pi \eta}{a^{2n+1}(n-m)!} \bigg[ \frac{4}{n^2 a} \left( B_{mn}^2 + \tilde{B}_{mn}^2\right) \nonumber \\
&+& \frac{n+2}{2n+1} \left( C_{mn}^2 + \tilde{C}_{mn}^2 \right)  \bigg]. \label{eqn:PowerTangential}
\end{eqnarray}
Note that the following identity 
\begin{eqnarray}
\int_{-1}^1 \left[ (1-\mu^2) P_n^{m'} P_l^{m'} +m^2 \frac{P_n^m P_l^m}{1-\mu^2} \right] d\mu = \frac{2n(n+1)(n+m)!}{(2n+1)(n-m)!} \delta_{nl} \label{eqn:Useful}, 
\end{eqnarray} 
has been derived and used in order to evaluate the necessary integrals in $\mathcal{P}$.

Using the result in Eq.~\ref{eqn:PowerTangential}, one can then compute the  hydrodynamic efficiency of a swimming squirmer, 
\begin{eqnarray}
\mathcal{E} = \frac{6 \pi \eta a \U^2}{\mathcal{P}},
\end{eqnarray}
and defined as the rate of work required to drag the spherical body at its swimming speed divided by the rate of work done by the self-propulsion to produce the same swimming speed \cite{Lighthill1975, Childress1981}.  {In the unsteady case, the efficiency is given by $6\pi \eta a \langle U \rangle^2 / \langle \mathcal{P} \rangle $, where $\langle ... \rangle$ denotes time-averaging.} In Eq.~\ref{eqn:PowerTangential} we see that all components contribute positive rate of work. However,   only the modes $B_{01}$, $B_{11}$, and $\tilde{B}_{11}$ contribute to the propulsion of the squirmer (see Eq.~\ref{eqn:NonAxisSpeed}). In other words, the inclusion of any other squirming modes leads to a less efficient swimmer. The same holds in the axisymmetric case, where the expression for the rate of work reduces to
\begin{eqnarray}
\mathcal{P} = \frac{64 \pi \eta}{3 a^5} B_{01}^2 + \sum_{n=2}^\infty \frac{4n(n+1) \pi \eta}{a^{2n+1}} \left( \frac{4}{n^2 a^2} B_{0n}^2 + \frac{n+2}{2n+1} C_{0n}^2\right), \label{eqn:PowerTangentialAxis}
\end{eqnarray}
a result which reduces to  Blake's \cite{Blake1971b} in the case of purely tangential deformation with the rescaling given by Eq.~\ref{eqn:rescale}. Note that the results for the axisymmetric case here, Eq.~\ref{eqn:PowerTangentialAxis}, are more general because of the inclusion of the rate of work by the azimuthal but axisymmetric components $C_{0n}$ that were not previously accounted for.

Interestingly, using the reciprocal theorem, Stone \& Samuel showed that the rate of work done by a general swimming organism is given by 
\begin{eqnarray}
\mathcal{P} = \eta \int_V \boldsymbol \omega^2 dV -2\eta \int_S \n \cdot (\v \cdot \nabla \v)dS, 
\end{eqnarray}
where $\boldsymbol \omega$ is the vorticity field in the fluid. Hence, for two swimmers propelling at the same speed, the one producing more vorticity dissipates more energy. It was therefore concluded that  it is less efficient for an axisymmetric object to rotate as its swims, compared with the corresponding non-rotating swimmer \cite{Stone1996}. 

In the explicit expression of the rate of work in the axisymmetric case (Eq.~\ref{eqn:PowerTangentialAxis}), the modes causing rotation of the squirmer $C_{11}$, $\tilde{C}_{11}$, and $C_{01}$ do not appear, which is not surprising given that these modes have no contribution to the flow field (see Eqs.~\ref{eqn:NonAxisSwimFlowR}--\ref{eqn:NonAxisSwimFlowPhi} and  Sec.~\ref{sec:AxisPhysics} for explanations), due to zero apparent rotation from the perspective of the fluid. In other words, a squirmer which self-rotates by exploiting purely the $C_{11}$, $\tilde{C}_{11}$, or $C_{01}$ modes induces no extra viscous dissipation. In this case, the swimmer  rotates as it swims but it  is as  efficient as a non-rotating swimmer, simply because the rotational motion alone does not produce any net flow (or vorticity). In other words,  no rotation is truly felt from the  perspective of the fluid, even though from the perspective of the swimmer itself there is a non-zero rotation rate.

In general however, a squirmer would have an azimuthal squirming profile containing not only the rotlet terms but also other modes $C_{mn}$ ($n \ge 2$), which would dissipate more energy, making the swimmer less efficient according to Eq.~\ref{eqn:PowerTangentialAxis}. The body rotation is apparent from the perspective of the fluid only when the azimuthal squirming profile contains modes $C_{mn}$ ($n \ge 2$) other than the rotlet terms, hence the conclusion by Stone \& Samuel \cite{Stone1996}.

\subsection{Squirming with arbitrary surface velocities}\label{sec:Arbitrary}
In previous sections, the surface velocities have been expressed in the form of the boundary values of Lamb's general solution, Eqs.~\ref{eqn:uthetaTanNonA}--\ref{eqn:uphiTanNonA}, and the flow structure under such a decomposition has been discussed. In general however, the surface velocities could be more naturally described using other surface decompositions. Here, we detail how to  relate surface velocities expressed in arbitrary forms to the decomposition employed in Lamb's general solution. 

For illustration, we decompose the arbitrary surface velocities in natural Fourier modes along the azimuthal direction $\phi$ as
\begin{eqnarray}
\u(a, \theta, \phi) =&& \ \sum_{m=0}^\infty \left[E_m (\theta) \cos m \phi + \tilde{E}_m (\theta) \sin m \phi\right] \e_\theta \nonumber \\
&+& \sum_{m=0}^\infty \left[F_m (\theta) \cos m \phi + \tilde{F}_m (\theta) \sin m \phi \right] \e_\phi \label{eqn:Tanproject1},
\end{eqnarray}
where $E_m(\theta), \tilde{E}_m(\theta), F_m(\theta), \tilde{F}_m(\theta)$ are  arbitrary functions in the polar direction projected from the boundary actuation under this decomposition. The goal here is therefore to derive the set of coefficients, $A_{mn}$, $\tilde{A}_{mn}$, $B_{mn}$, $\tilde{B}_{mn}$, $C_{mn}$, $\tilde{C}_{mn}$,  in Lamb's general solution, Eqs.~\ref{eqn:uthetaTanNonA}--\ref{eqn:uphiTanNonA},  given the surface velocities expressed in Eq.~\ref{eqn:Tanproject1}.

The attempt of directly projecting Eq.~\ref{eqn:Tanproject1} onto Eqs.~\ref{eqn:uthetaTanNonA}--\ref{eqn:uphiTanNonA} in order to calculate the coefficients is non-trivial, because a mix of basis functions is used in Eq.~\ref{eqn:uthetaTanNonA}--\ref{eqn:uphiTanNonA} and there is no obvious way of doing orthogonal projections to calculate the coefficients (unless $m=0$ for the axisymmetric case). We instead follow a systematic scheme due to Brenner \cite{Brenner1964}  facilitating the projections of boundary condition when using Lamb's general solution. {While the radial velocity on the sphere is still matched, in place of matching the polar and azimuthal velocity components on the sphere, the quantities $r (\nabla \cdot \u|_{r=a})$ and $r\e_r \cdot (\nabla \times \u|_{r=a})$ are matched \cite{Brenner1964, Happel1973, Kim2004}, leading to}
\begin{eqnarray}
\e_r \cdot \u|_{r=a} =&& \sum_{n=1}^\infty \bigg[ \frac{(n+1)a p_{-(n+1)}|_{r=a}}{2\eta (2n-1)} \nonumber \\
&-&\frac{n+1}{a} \Phi_{-(n+1)}\big|_{r=a}   \bigg], \label{eqn:matching1}\\
- r \nabla \cdot (\u|_{r=a}) =&& \sum_{n=1}^\infty \bigg[  -\frac{n (n+1)a}{2\eta (2n-1)} p_{-(n+1)}|_{r=a} \nonumber \\
&+& \frac{(n+1)(n+2)}{a} \Phi_{-(n+1)}\big|_{r=a} \bigg], \label{eqn:matching2}\\
r\e_r \cdot (\nabla \times \u|_{r=a}) &=& \sum_{n=1}^\infty n(n+1)\chi_{-(n+1)}\big|_{r=a}, \label{eqn:matching3}
\end{eqnarray}
in terms of Lamb's general solution.

In the case of the purely tangential deformation, the first matching condition, Eq.~\ref{eqn:matching1}, is simply the conditions for no radial deformation, Eq.~\ref{eqn:purelyTangential}, relating $B_{mn}$ and $\tilde{B}_{mn}$ to $A_{mn}$ and $\tilde{A}_{mn}$. We are therefore only left with Eqs.~\ref{eqn:matching2} and \ref{eqn:matching3} to determine the remaining coefficients. Expressed in spherical coordinates, Eqs.~\ref{eqn:matching2} and \ref{eqn:matching3} are
\begin{eqnarray}
&-&2 u_r|_{r=a} - \frac{1}{\sin \theta} \bigg[ \frac{\partial}{\partial \theta} \left(u_\theta\sin\theta \right) + \frac{\partial u_\phi}{\partial \phi} \bigg]_{r=a} \nonumber \\
 &=& \sum_{n=1}^\infty \sum_{m=0}^n  \frac{2(n+1)}{a^{n+2}}P_n^m \left(B_{mn}\cos m\phi +\tilde{B}_{mn} \sin m\phi \right),\label{eqn:Brenner1}\\
&&\frac{1}{\sin \theta} \left[ \frac{\partial}{\partial \theta} \left(u_\phi \sin \theta\right)-\frac{\partial u_\theta}{\partial \phi} \right]_{r=a} \nonumber \\
&=& \sum_{n=1}^\infty \sum_{m=0}^n \frac{n(n+1)}{a^{n+1}}  P_n^m \left(C_{mn} \cos m\phi+\tilde{C}_{mn}\sin m \phi \right) . \label{eqn:Brenner2}
\end{eqnarray}

The benefits of Brenner's matching conditions are now clear because the coefficients on the right hand side of Eqs.~\ref{eqn:Brenner1}--\ref{eqn:Brenner2} can be readily determined by the orthogonality of associated Legendre polynomials. Substituting the prescribed boundary conditions, Eq.~\ref{eqn:Tanproject1}, into the above matching conditions yields the two  equations
\begin{eqnarray}
&-& \frac{1}{\sin \theta} \frac{\partial}{\partial \theta} \left(E_0 \sin \theta\right) -\sum_{m=1}^\infty \Bigg\{ \left[\frac{1}{\sin \theta} \frac{\partial}{\partial \theta} \left(E_m \sin \theta\right) + m \frac{\tilde{F}_m}{\sin \theta}  \right] \cos m\phi  \nonumber \\
&+&\left[ \frac{1}{\sin \theta} \frac{\partial}{\partial \theta} \left(\tilde{E}_m \sin \theta\right) + m \frac{F_m}{\sin \theta}  \right] \sin m\phi \Bigg\} \nonumber \\
=&&\sum_{n=1}^\infty  \frac{2(n+1)}{a^{n+2}}P_n B_{0n} \nonumber \\
&+& \sum_{n=1}^\infty \sum_{n=m}^\infty  \frac{2(n+1)}{a^{n+2}}P_n^m \left(B_{mn}\cos m\phi +\tilde{B}_{mn} \sin m\phi \right), 
\end{eqnarray}
\begin{eqnarray}
&&\frac{1}{\sin \theta} \frac{\partial}{\partial \theta} \left( F_0 \sin \theta \right) + \sum_{n=1}^\infty  \frac{1}{\sin \theta} \Bigg\{ \left[ \frac{\partial}{\partial \theta} \left( F_m \sin \theta \right)  -m \tilde{E}_m\right] \cos m\phi \nonumber \\
&+&\left[ \frac{\partial}{\partial \theta} \left( \tilde{F}_m \sin \theta\right)  +m E_m  \right] \sin m\phi \Bigg\} \nonumber \\
=&& \sum_{n=1}^\infty \frac{n (n+1)}{a^{n+1}} P_n C_{0n} \nonumber \\
&+& \sum_{m=1}^\infty \sum_{n=m}^\infty \frac{n(n+1)}{a^{n+1}} P_n^m \left( C_{mn} \cos m\phi+ \tilde{C}_{mn} \sin m\phi \right) .
\end{eqnarray}

Finally, by the orthogonality of associated Legendre polynomials we obtain the remaining explicit expressions for the Lamb coefficients as
\begin{eqnarray}
B_{mn} &=& \frac{a^{n+2}}{4} \frac{(2n+1)(n-m)!}{(n+1)(n+m)!} \times \nonumber\\
&& \int_{-1}^1 \left[ \frac{\partial}{\partial \mu} \left(E_m \sin \theta \right) -\frac{m \tilde{F}_m}{\sin \theta} \right] P_n^m d\mu, \label{eqn:ProjectResultsTan1}\\
\tilde{B}_{mn} &=& \frac{a^{n+2}}{4} \frac{(2n+1)(n-m)!}{(n+1)(n+m)!} \times \nonumber \\
&& \int_{-1}^1  \left[ \frac{\partial}{\partial \mu} \left( \tilde{E}_m \sin \theta\right)+ \frac{m F_m}{\sin \theta} \right] P_n^m d\mu,\\
C_{mn} &=& \frac{a^{n+1}}{2n}\frac{(2n+1)(n-m)!}{(n+1)(n+m)!} \times \nonumber \\
&&\int_{-1}^1 \left[ -\frac{\partial}{\partial \mu} \left( F_m \sin \theta \right) -\frac{m \tilde{E}_m}{\sin \theta} \right] P_n^m d\mu, \\
\tilde{C}_{mn} &=& \frac{a^{n+1}}{2n}\frac{(2n+1)(n-m)!}{(n+1)(n+m)!} \times \nonumber \\
&&\int_{-1}^1 \left[ -\frac{\partial}{\partial \mu} \left( \tilde{F}_m \sin \theta \right) +\frac{m E_m}{\sin \theta} \right] P_n^m d\mu, \label{eqn:ProjectResultsTan2}
\end{eqnarray}
for $0\le m \le \infty$ and $m \le n \neq0$. Note that the coefficients $A_{mn}$ and $\tilde{A}_{mn}$ are given by the no radial deformation conditions, Eq.~\ref{eqn:purelyTangential}. The more general analysis including  radial deformation at the squirmer surface is presented in Appendix \ref{app:AppendixGeneral}.

\section{Discussion}\label{sec:Discuss}
In this paper, we have investigated the squirming motion of a sphere using Lamb's general solution. With this alternative formulation, our results first complete the classical analysis of axisymmetric squirming motion by including the azimuthal velocity fields not taken into account in previous studies (Sec.~\ref{sec:Axis}). We then  extended the analysis to the general non-axisymmetric case in order to study the motion of a   squirmer performing arbitrary  three-dimensional translation and rotation (Sec.~\ref{sec:NonAxisMain}). Analytical formulae for both  the swimming kinematics and the complete flow field are derived (Sec.~\ref{sec:NonAxis}). In the axisymmetric case, the motion of a squirmer is restricted to a straight line and  removing the axisymmetry frees the squirmer to move with helical trajectories in general, with circular and straight paths as special limits.

As summarized in Sec.~\ref{sec:reciprocal}, the swimming kinematics of arbitrary squirming profiles can  be obtained using a reciprocal theorem approach. However, that approach does not allow to gain  information on the flow induced by the squirmer, which is important for the computation of the swimmer hydrodynamic efficiency and for problems such as the transport and uptake of nutrients of microorganisms. In our paper, we obtained the flow field around the squirmer  analytically   using  Lamb's general solution, and we interpreted the flow structure   in terms of fundamental singularities in Stokes flows (Sec.~\ref{sec:NonAxisPhysics}). We have also remarked on the power dissipation and efficiency of a generalized squirmer (Sec.~\ref{sec:Rate}), and detailed the general procedure of relating surface velocities of arbitrary forms to the form used in Lamb's general solution (Sec.~\ref{sec:Arbitrary}). 

The traditional, axisymmetric squirmer model has been widely adopted to describe pushers and pullers, by changing the sign of the stresslet component relative to the source dipole ($B_{02}/B_{01}$), thereby allowing to address many problems in biological physics (as detailed in the introduction). With the axisymmetric azimuthal modes identified in this work, the squirmer model may, for instance, be modified to incorporate a rotlet dipole to represent the effect of the rotating cell body and the counter-rotating flagellum of \textit{Escherichia coli}. The freedom to choose the sign of the rotlet dipole relative to the source dipole ($C_{02}/B_{01}$) may also be useful in the study of the switching of rotation direction in bacterial flagella during tumbling processes \cite{Turner2000}.  Higher order, non-axisymmetric, modes may now also be included to model   more complex hydrodynamic effects. This generalized squirmer model may then be useful for investigating three-dimensional  swimmer-swimmer, or swimmer-boundary interactions.

\section*{Acknowledgements}
Funding from the US National Science Foundation (Grant CBET-0746285 to E.~L.) and the Croucher Foundation (through a fellowship to O.~S.~P.) is gratefully acknowledged. The authors also wish to thank the  Department of Mechanical and Aerospace Engineering at the University of California, San Diego where this work was initiated.

\begin{appendix} \label{sec:Appendix}
\section{Fundamental flow singularities in spherical coordinates}\label{app.singularities}

\subsection{Stokeslets}\label{app:Stokeslet}
{The solution to Stokes equation due to a point force $f \alphaB \delta(\r)$ of magnitude $f$ and direction $\alphaB$ at the origin is given by $\u = f \G(\alphaB)/ (8\pi \eta)$, where the vectorial representation of a Stokeslet is given by Eq.~\ref{eqn:Stokeslet} in the main text.}

Note that $\alphaB$ denotes the direction of the Stokeslet.  Stokeslets in different Cartesian directions are expressed in spherical coordinates as
\begin{eqnarray}
\G(\e_x) &=& \frac{1}{r} \left[  2 \sin \theta \cos \phi \ \e_r + \cos \theta \cos \phi \ \e_\theta -\sin \phi \ \e_\phi \right], \\
\G(\e_y) &=&  \frac{1}{r} \left[  2 \sin \theta \sin \phi \ \e_r + \cos \theta \sin \phi \ \e_\theta + \cos \phi \ \e_\phi \right],\\
\G(\e_z) &=& \frac{1}{r} \left[  2 \cos \theta \ \e_r - \sin \theta \ \e_\theta  \right] .
\end{eqnarray}

\subsection{A general Stokes dipole}\label{app:StokesDipole}
A general force dipole is obtained by taking the derivative of a Stokeslet along the direction of interest. The vectorial representation of a force dipole is given by Eq.~\ref{eqn:SD} in the main text,
where $\alphaB$ and $\betaB$ denote the direction of the Stokeslet and the direction along which the derivative is taken respectively. The expressions of Stokes dipoles of different configurations in spherical coordinates are given by
\begin{eqnarray}
\G_D(\e_x, \e_x) =& \frac{1}{r^2} \left[ -\frac{1}{4}(1+3 \cos 2\theta) + \frac{3}{4} (1-\cos2\theta) \cos 2\phi \right] \e_r,  \\
\G_D(\e_y, \e_y) =& \frac{1}{r^2} \left[ -\frac{1}{4} (1+3 \cos 2\theta)  -\frac{3}{4} (1-\cos2\theta) \cos 2\phi \right] \e_r, \\
\G_D(\e_z, \e_z) =& \frac{1}{2r^2} (1+3 \cos 2\theta) \e_r, \\
\G_D(\e_y, \e_x) =&  \frac{1}{r^2} \left[ \frac{3}{4} (1-\cos2\theta) \sin 2\phi \ \e_r  - \sin \theta \ \e_\phi \right],\\
\G_D(\e_z, \e_x) =& \frac{1}{r^2} \left( \frac{3}{2} \sin2\theta \cos \phi \ \e_r + \cos \phi \ \e_\theta - \cos \theta \sin \phi \ \e_\phi \right), \\
\G_D(\e_x, \e_y) =& \frac{1}{r^2} \left[ \frac{3}{4} (1-\cos2\theta)  \sin 2\phi \ \e_r + \sin\theta \ \e_\phi  \right],\\
\G_D(\e_z, \e_y) =& \frac{1}{r^2} \left( \frac{3}{2} \sin 2\theta \sin \phi \ \e_r + \sin \phi \ \e_\theta + \cos \theta \cos\phi \ \e_\phi \right),\\
\G_D(\e_x, \e_z) =&  \frac{1}{r^2}\left( \frac{3}{2} \sin2\theta \cos \phi \ \e_r -\cos\phi \ \e_\theta + \cos \theta \sin \phi \ \e_\phi \right),\\
\G_D(\e_y, \e_z)  =& \frac{1}{r^2} \left( \frac{3}{2} \sin2\theta \sin \phi \ \e_r - \sin \phi \ \e_\theta - \cos\theta \cos \phi \ \e_\phi \right) .
\end{eqnarray}

\subsection{Stresslets}\label{app:AppendixStresslet}
The vectorial representation of a stresslet is given by Eq.~\ref{eqn:DefStresslet} in the main text,
where $\alphaB$ and $\betaB$ denote the direction of the Stokeslet and the direction along which the derivative is taken respectively. The expressions of stresslets of different configurations in spherical coordinates are given by
\begin{eqnarray}
\SS(\e_x, \e_x) &=& \G_D(\e_x, \e_x) \nonumber \\
&=& \frac{1}{r^2} \left[ -\frac{1}{4}(1+3 \cos 2\theta) + \frac{3}{4} (1-\cos2\theta) \cos 2\phi \right] \e_r,  \\
\SS(\e_y, \e_y) &=&\G_D(\e_y \e_y) \nonumber \\
&=& \frac{1}{r^2} \left[ -\frac{1}{4} (1+3 \cos 2\theta)  -\frac{3}{4} (1-\cos2\theta) \cos 2\phi \right] \e_r, \\
\SS(\e_z, \e_z) &=&\G_D(\e_z,\e_z) = \frac{1}{2r^2} (1+3 \cos 2\theta) \e_r, \\
\SS(\e_y, \e_x) &=&\SS(\e_x, \e_y)=   \frac{3}{4r^2} (1-\cos2\theta) \sin 2\phi \ \e_r  ,\\
\SS(\e_z, \e_x) &=&\SS(\e_x, \e_z)= \frac{3}{2r^2} \sin2\theta \cos \phi \ \e_r , \\
\SS(\e_z, \e_y) &=& \SS(\e_y, \e_z) =  \frac{3}{2r^2} \sin 2\theta \sin \phi \ \e_r .
\end{eqnarray}

\subsection{Rotlets}\label{app:AppendixRotlet}
The vectorial representation of a rotlet is given by Eq.~\ref{eqn:rotlet} in the main text,
where $\zetaB = \betaB \times \alphaB$ denotes the direction of the rotlet. The expressions of rotlets in different directions in spherical coordinates are given by 
\begin{eqnarray}
\R(\e_x) &=&  \frac{1}{r^2}\left( - \sin \phi \ \e_\theta -\cos \theta \cos \phi \ \e_\phi \right), \\
\R(\e_y) &=&  \frac{ 1}{r^2} \left(\cos \phi \ \e_\theta -\cos \theta \sin \phi \ \e_\phi \right),\\
\R(\e_z) &=& \frac{1}{r^2} \left( \sin \theta \ \e_\phi\right) .
\end{eqnarray}

\subsection{A general Stokes quadrupole}\label{app:quadrupole}
A higher-order singularity, the force quadrupole, is obtained by taking the derivative of a force dipole along different directions and is given by Eq.~\ref{eqn:GSQ} in the main text,
where $\betaB, \gammaB$ are the directions along which each derivative is taken. The expressions of Stokes quadrupoles of different configurations in spherical coordinates are given by
\begin{eqnarray}
\G(\e_x,\e_x,\e_x) =&& \ \frac{1}{4r^3} \left[ - (5 \sin\theta+9\sin3\theta)\cos\phi + 3 (3\sin\theta-\sin3\theta)\cos3\phi \right] \e_r \nonumber \\
&+& \frac{1}{16r^3} \left[(7 \cos \theta+9\cos3\theta)\cos \phi-3 (\cos\theta-\cos3\theta) \cos3\phi \right] \e_\theta \nonumber \\
&+& \frac{1}{8r^3} \left[-(5+3\cos2\theta)\sin\phi + 3(1-\cos2\theta)\sin3\phi \right] \e_\phi,
\end{eqnarray}
\begin{eqnarray}
\G(\e_y,\e_x,\e_x)=&&\G(\e_x,\e_y,\e_x) \nonumber \\
=&& \ \frac{1}{4r^3} \left[ (\sin\theta-3\sin3\theta)\sin\phi+3(3\sin\theta-\sin3\theta) \sin3\phi \right] \e_r \nonumber\\
&+& \frac{1}{16r^3} \left[ (13 \cos\theta+3\cos3\theta)\sin\phi+ 3(\cos3\theta-\cos\theta)\sin3\phi \right] \e_\theta \nonumber \\
&+& \frac{1}{8r^3} \left[ (9\cos2\theta-1)\cos\phi-3(1-\cos2\theta)  \cos3\phi \right] \e_\phi, 
\end{eqnarray}
\begin{eqnarray}
\G(\e_z, \e_x,\e_x) =&&\G(\e_x, \e_z,\e_x) \nonumber \\
=&& \ \frac{1}{2r^3} \left[ - (\cos\theta+3\cos3\theta)+3 (\cos\theta-\cos3\theta)\cos2\phi \right] \e_r \nonumber \\
&+& \frac{1}{8r^3} \times \nonumber \\
&& \left[ (\sin\theta-3\sin3\theta)+3 (3\sin\theta-\sin3\theta)\cos2\phi\right] \e_\theta, 
\end{eqnarray}
\begin{eqnarray}
\G(e_y,\e_y,\e_x) =&& \ \frac{1}{4r^3} \left[ - (7 \sin\theta+3\sin3\theta)\cos\phi-3 (3\sin\theta-\sin3\theta)\cos3\phi \right] \e_r \nonumber \\
&+& \frac{1}{16r^3} \left[(3\cos3\theta-19\cos\theta)\cos\phi + 3(\cos\theta-\cos3\theta)\cos3\phi \right] \e_\theta \nonumber \\
&+& \frac{1}{8r^3} \left[ (15\cos2\theta-7)\sin\phi-3(1-\cos2\theta) \sin3\phi \right] \e_\phi,
\end{eqnarray}
\begin{eqnarray}
\G(\e_z,\e_y,\e_x) =&& \ \G(\e_y,\e_z,\e_x) \nonumber \\
=&& \ \frac{1}{r^3} \bigg[ \frac{3}{2}(\cos\theta-\cos3\theta)\sin2\phi \ \e_r \nonumber \\
&+& \frac{3}{8} (3\sin\theta-\sin3\theta)\sin2\phi \ \e_\theta-\frac{3}{2}\sin2\theta \ \e_\phi \bigg], 
\end{eqnarray}
\begin{eqnarray}
\G(\e_z,\e_z,\e_x) =&& \ \frac{1}{r^3} \bigg[ (3\sin3\theta-\sin\theta) \cos\phi  \ \e_r + \frac{1}{4} (11\cos\theta-3\cos3\theta)\cos\phi \ \e_\theta \nonumber \\
&-& \frac{1}{2}(1+3\cos2\theta)\sin \phi  \ \e_\phi \bigg],
\end{eqnarray}
\begin{eqnarray}
\G(\e_x,\e_x,\e_y) =& & \frac{1}{4r^3} \left[ -(7 \sin\theta + 3\sin3\theta) \sin\phi+3 (3\sin\theta-\sin3\theta) \sin3\phi \right] \e_r \nonumber\\
&+& \frac{1}{16r^3} \left[ \cos\theta (-19+3\cos2\theta)\sin\phi-3(\cos\theta-\cos3\theta)\sin3\phi \right]\e_\theta \nonumber \\
&+& \frac{1}{8r^3} \left[ (7-15 \cos2\theta)\cos\phi - 3(1-\cos2\theta)\cos3\phi\right]\e_\phi,
\end{eqnarray}
\begin{eqnarray}
\G(\e_y, \e_x, \e_y) =&&\G(\e_x, \e_y, \e_y) \nonumber \\
=&& \ \frac{1}{4r^3} \left[ (\sin\theta-3\sin3\theta) \cos\phi- 3(3\sin\theta-\sin3\theta) \cos3\phi \right]\e_r \nonumber \\
&+& \frac{1}{16r^3} \left[ (13 \cos\theta+3\cos3\theta)\cos\phi+3(\cos\theta-\cos3\theta)\cos3\phi \right] \e_\theta \nonumber \\
&+& \frac{1}{8r^3} \left[ (1-9\cos2\theta)\sin\phi-3(1-\cos2\theta)  \sin3\phi  \right] \e_\phi, 
\end{eqnarray}
\begin{eqnarray}
\G(\e_z,\e_x,\e_y) =&& \G(\e_x,\e_z,\e_y) \nonumber \\
 =&& \frac{1}{r^3} \bigg[ \frac{3}{2} (\cos\theta-\cos3\theta)\sin2\phi \ \e_r  \nonumber \\
 &+& \frac{3}{8}(3\sin\theta-\sin3\theta)\sin2\phi \ \e_\theta + \frac{3}{2}\sin2\theta \ \e_\phi \bigg],
\end{eqnarray}
\begin{eqnarray}
\G(\e_y,\e_y,\e_y) =&& \ \frac{1}{4r^3} \left[ -(5\sin\theta+9\sin3\theta)\sin\phi-3(3\sin\theta-\sin3\theta)\sin3\phi  \right] \e_r \nonumber \\
&+& \frac{1}{16r^3} \left[ (7\cos\theta+9\cos3\theta) \sin\phi+3(\cos\theta-\cos3\theta)\sin3\phi \right] \e_\theta \nonumber\\
&+& \frac{1}{8r^3} \left[ (5+3\cos2\theta)\cos\phi + 3 (1-\cos2\theta)  \cos3\phi \right] \e_\phi, 
\end{eqnarray}
\begin{eqnarray}
\G(\e_z, \e_y, \e_y)=&& \G(\e_y, \e_z, \e_y) \nonumber  \\
=&& \ \frac{1}{2r^3} \left[ - (\cos\theta+3\cos3\theta)- 3(\cos\theta-\cos3\theta) \cos2\phi  \right] \e_r \nonumber \\
&+& \frac{1}{8r^3} \times \nonumber \\
&& \left[  (\sin\theta-3\sin3\theta)-3 (3\sin \theta-\sin 3 \theta)\cos2\phi  \right] \e_\theta, 
\end{eqnarray}
\begin{eqnarray}
\G(\e_z,\e_z,\e_y) =&& \frac{1}{r^3} \bigg[ (3\sin\theta-\sin\theta)\sin\phi \ \e_r+ \frac{1}{4}(11\cos\theta-3\cos3\theta)\sin\phi \ \e_\theta \nonumber \\
&+& \frac{1}{2}(1+3\cos2\theta)\cos\phi \ \e_\phi   \bigg], 
\end{eqnarray}
\begin{eqnarray}
\G(\e_x,\e_x,\e_z) =&&  \frac{1}{2r^3}\left[ -(5 \cos \theta+3 \cos3\theta)+ 3(\cos\theta-\cos3\theta)\cos2\phi  \right] \e_r \nonumber \\
&-& \frac{1}{8r^3} \left[ (7\sin\theta+3\sin3\theta)+ 3(5\sin\theta+\sin3\theta)\cos2\phi\right] \e_\theta \nonumber \\
&+& \frac{3}{2r^3} \sin2\theta \sin 2\phi \ \e_\phi, 
\end{eqnarray}
\begin{eqnarray}
\G(\e_y, \e_x, \e_z)=&& \G(\e_x,\e_y,\e_z) \nonumber \\
=&&  \frac{1}{r^3} \bigg[ \frac{3}{2}(\cos\theta-\cos3\theta)\sin2\phi \ \e_r - \frac{3}{8} (5\sin\theta+\sin3\theta) \sin2\phi \ \e_\theta \nonumber \\
&-& \frac{3}{2} \sin2\theta\cos2\phi \ \e_\phi \bigg], 
\end{eqnarray}
\begin{eqnarray}
\G(\e_z,\e_x,\e_z) =&& \G(\e_x,\e_z,\e_z) \nonumber \\
=&&  \frac{1}{r^3} \bigg[ (\sin\theta+3\sin3\theta)\cos\phi \ \e_r- \frac{1}{4}(5\cos\theta+3\cos3\theta)\cos\phi \ \e_\theta \nonumber \\
&+& \frac{1}{2}(1+3\cos2\theta) \ \e_\phi \bigg], 
\end{eqnarray}
\begin{eqnarray}
\G(\e_y,\e_y,\e_z) =&& \  \frac{1}{2r^3} \left[ -(5\cos\theta+3\cos3\theta) - 3(\cos\theta-\cos3\theta) \cos2\phi  \right] \e_r \nonumber \\
&+& \frac{1}{8r^3} \left[ -(7\sin\theta+3\sin3\theta) + 3(5\sin\theta+\sin3\theta)\cos2\phi  \right] \e_\theta  \nonumber \\
&-& \frac{3}{2r^3} \sin2\theta \sin2\phi \ \e_\phi,
\end{eqnarray}
\begin{eqnarray}
\G(\e_z, \e_y, \e_z) =&&\G(\e_y, \e_z, \e_z) \nonumber \\
=&& \frac{1}{r^3} \bigg[ (\sin\theta+3\sin3\theta)\sin\phi  \ \e_r  - \frac{1}{4} (5\cos\theta+3\cos3\theta)\sin\phi \ \e_\theta \nonumber \\
&-& \frac{1}{2} (1+3\cos2\theta) \cos\phi \ \e_\phi \bigg], 
\end{eqnarray}
\begin{eqnarray}
\G(\e_z,\e_z,\e_z) = \frac{1}{r^3} \bigg[ (\cos\theta+3\cos3\theta) \ \e_r  + \frac{1}{4}(3\sin3\theta-\sin\theta) \ \e_\theta \bigg] .
\end{eqnarray}

\subsection{Potential Dipoles}\label{app:PotentialDipole}
The vectorial representation of a potential (source) dipole is given by Eq.~\ref{eqn:podi} in the main text,
where $\alphaB$ denotes the direction of the  dipole. The expressions of potential dipoles in different directions in spherical coordinates are given by 
\begin{eqnarray}
\P_D(\e_x) &=& \frac{1}{r^3} \left[  2 \sin \theta \cos \phi \ \e_r  -\cos \theta \cos \phi \ \e_\theta +  \sin \phi \ \e_\phi \right], \\
\P_D(\e_y) &=&  \frac{1}{r^3} \left[ 2 \sin \theta \sin \phi \ \e_r - \cos \theta \sin \phi \  \e_\theta - \cos \phi \ \e_\phi \right],\\
\P_D(\e_z) &=& \frac{1}{r^3} \left[ 2 \cos \theta \ \e_r + \sin \theta \ \e_\theta \right] .
\end{eqnarray}

\subsection{Rotlet dipoles}\label{app:RotletDipole}
One can take a derivative of a rotlet to obtain a rotlet dipole, which is given by Eq.~\ref{eqb:RotDi} in the main text, 
where $\zetaB$ and $\gammaB$ denote the direction of the rotlet and the direction along which the derivative is taken respectively. The expressions of rotlet dipoles of different configurations in spherical coordinates are given by
\begin{eqnarray}
\R_D(\e_x, \e_x) =&&  \frac{1}{r^3} \bigg[ -\frac{3}{2} \sin \theta \sin 2 \phi \ \e_\theta \nonumber \\
&-& \frac{3}{4} \sin2\theta \left(1+\cos 2\phi \right) \e_\phi \bigg], \\
\R_D(\e_y,\e_y) =&&  \frac{1}{r^3} \bigg[ \frac{3}{2} \sin \theta \sin 2 \phi \ \e_\theta \nonumber\\
&-& \frac{3}{4} \sin 2 \theta \left(1- \cos 2\phi \right) \e_\phi   \bigg],\\
\R_D(\e_z,\e_z) =&& \frac{3 \sin 2\theta}{2r^3}\e_\phi, \\
\R_D(\e_y,\e_x) =&&  \frac{1}{r^3} \bigg[ -\cos \theta \ \e_r -\frac{1}{2}\sin\theta\left(1-3 \cos 2\phi \right) \ \e_\theta \nonumber \\
&-& \frac{3}{4} \sin 2\theta \sin2 \phi \ \e_\phi \bigg],\\
\R_D(\e_z,\e_x)=&& \frac{1}{r^3} \bigg[ \sin \theta \sin \phi \ \e_r  - 2 \cos \theta \sin \phi  \ \e_\theta \nonumber \\
&-&\frac{1}{2} \left(1+3 \cos 2\theta \right) \cos \phi \ \e_\phi \bigg],\\
\R_D(\e_x,\e_y)=&& \frac{1}{r^3} \bigg[ \cos \theta \ \e_r + \frac{1}{2}\sin\theta\left(1 +3 \cos 2\phi\right) \ \e_\theta \nonumber \\ 
&-& \frac{3}{4} \sin2 \theta \sin2 \phi \ \e_\phi  \bigg], \\
\R_D(\e_z,\e_y)=&& \frac{1}{r^3} \bigg[ -\sin \theta \cos \phi \ \e_r + 2 \cos \theta \cos \phi \ \e_\theta \nonumber \\ 
&-&\frac{1}{2}\left(1+3 \cos 2\theta \right) \sin \phi \ \e_\phi   \bigg], \\
\R_D(\e_x,\e_z)=&&\frac{1}{r^3} \bigg[ -\sin \theta \sin \phi \ \e_r  -\cos \theta \sin \phi  \ \e_\theta  \nonumber \\
&+&\frac{1}{2} \left(1-3\cos2\theta  \right) \cos \phi \ \e_\phi   \bigg], \\
\R_D(\e_y,\e_z)=&& \frac{1}{r^3} \bigg[ \sin \theta \cos \phi \ \e_r + \cos \theta \cos \phi \ \e_\theta \nonumber \\
&+&\frac{1}{2}\left(1-3\cos 2\theta\right) \sin \phi \ \e_\phi \bigg] .
\end{eqnarray}

\section{Swimming of a squirmer with radial deformation}\label{app:AppendixSwimming}

{In the main text, we considered squirmers with purely tangential deformation. In this appendix, we complement these results  by addressing the case of squirmers  also undergoing radial deformation. The results here might also be useful for modeling jet-driven microscopic swimmers, for instance, the locomotion of bacteria expelling slime.}

For the axisymmetric case without the restriction to purely tangential deformation, Eq.~\ref{eqn:purelyTangential}, the solution to the pumping problem reads
\begin{eqnarray}
u_r   &=&\sum_{n=1}^\infty \frac{(n+1) P_n}{2 (2n-1) \eta r^{n+2}} \left[ A_{0n} r^2 - 2 B_{0n} (2n-1) \eta \right],\\
u_\theta  &=& \sum_{n=1}^\infty  \frac{\sin \theta P_n^{'}}{2 r^{n}} \left[  \frac{ n-2}{n(2n-1) \eta} A_{0n}    - \frac{2}{r^2} B_{0n}  \right], \\
u_\phi &=&\sum_{n=1}^\infty \frac{\sin\theta P_n^{'}}{r^{n+1}}  C_{0n},
\end{eqnarray}
with  the surface velocities
\begin{eqnarray}
u_r(r=a)   &=&\sum_{n=1}^\infty \left[\frac{(n+1)A_{0n} }{2 (2n-1) \eta a^{n}}  - \frac{B_{0n}}{a^{n+2}}  \right] P_n,\\
u_\theta (r=a)  &=& \sum_{n=1}^\infty\left[  \frac{ n-2}{2n(2n-1)a^n \eta} A_{0n}    - \frac{1}{a^{n+2}} B_{0n}  \right] \sin \theta P_n^{'} , \\
u_\phi (r=a) &=&\sum_{n=1}^\infty \frac{\sin\theta P_n^{'}}{a^{n+1}}  C_{0n} .
\end{eqnarray}
Without the azimuthal modes $C_{0n}$, the above surface velocities reduce to the form in Lighthill \cite{Lighthill1952} and Blake \cite{Blake1971b}. However, notice that the coefficients $A_{0n}$ and $B_{0n}$ do not correspond directly to the coefficients $A_n$ and $B_n$ used in Lighthill \cite{Lighthill1952} and Blake \cite{Blake1971b}, which represent directly the radial and polar modes respectively. With Lamb's general solution, the radial and polar modes are represented by a combination of the $A_{0n}$ and $B_{0n}$ modes. The relation between the two sets of coefficients is given by
\begin{eqnarray}
A_{0n} &=& \frac{a^n n (2n-1)\eta}{n+1} A_n - \frac{2a^n (2n-1)\eta}{n+1} B_n, \label{eqn:PakLB1} \\
B_{0n} &=& \frac{a^{n+2}(n-2)}{2(n+1)}A_n - \frac{a^{n+2}}{n+1}B_n . \label{eqn:PakLB2}
\end{eqnarray}

The translational and rotational velocities are computed similarly to Sec.~\ref{sec:AxisSwimming}. Without the restriction to tangential deformation, Eq.~\ref{eqn:purelyTangential}, the propulsion velocity becomes
\begin{eqnarray}
\U &=& \frac{2}{3 \eta a} \nabla\left[ r \left( P_1 A_{01} \right)  \right] = -\frac{2}{3a \eta} A_{01} \e_z,
\end{eqnarray}
while the computation of the rotational velocity is unaffected (Eq.~\ref{eqn:AxisRotation}).

The flow field around an axisymmetric swimming squirmer is thus given by 
\begin{eqnarray}
v_r  =&&  \left( \frac{a^2}{3 \eta}A_{01}-2B_{01} \right) \frac{\cos \theta}{r^3} \nonumber \\
&+& \sum_{n=2}^\infty  \frac{(n+1) P_n}{r^{n+2}} \left[ \frac{A_{0n} r^2}{2(2n-1) \eta} - B_{0n} \right]  , \\
v_\theta =&& \left(\frac{a^2}{6\eta} A_{01}-B_{01}\right) \frac{\sin\theta}{r^3} \nonumber \\
&+& \sum_{n=2}^\infty \frac{\sin \theta P_n^{'}}{r^{n+2}} \left[ \frac{(n-2)r^2}{2n (2n-1) \eta} A_{0n} - B_{0n}\right], \\
v_\phi =&& \sum_{n=2}^\infty  \frac{\sin \theta P_n^{'}}{r^{n+1}} C_{0n} .
\end{eqnarray}

The computation of the propulsion speed of a non-axisymmetric squirmer with radial deformation follows the same procedures in Sec.~\ref{sec:NonAxis}, but without the restriction to tangential deformation, Eq.~\ref{eqn:purelyTangential}. The propulsion speed is given by
\begin{eqnarray}
\U &=& \frac{2}{3 \eta a} \nabla\left[ r \left( P_1 A_{01}+ P_1^1 \cos \phi A_{11}+ P_1^1 \sin \phi \tilde{A}_{11} \right)  \right] \nonumber \\
&=& \frac{2}{3a \eta} \left( A_{11} \e_x + \tilde{A}_{11} \e_y - A_{01} \e_z \right),
\end{eqnarray}
while the expression for the rotational speed remains the same, Eq.~\ref{eqn:NonAxisRotSpeed}.

{To obtain the overall swimming flow field, we follow the same procedures as in the axisymmetric case.} Superimposing Lamb's general solution in the pumping problem, Eqs.~\ref{eqn:ur}--\ref{eqn:uphi}, with the flow fields due to the induced translation and rotation at the velocities determined above, we arrive at the flow field surrounding a general swimming squirmer without the assumption of purely tangential deformation
\begin{eqnarray}
v_r  =&&  \frac{1}{r^3} \bigg[ \left(2 B_{11} - \frac{a^2}{3 \eta}A_{11} \right)\sin \theta \cos\phi  +  \left(2 \tilde{B}_{11}- \frac{a^2}{3 \eta} \tilde{A}_{11} \right)\sin \theta \sin \phi \nonumber \\
&-&  \left( 2B_{01}- \frac{a^2}{3 \eta}A_{01} \right) \cos \theta \bigg] + \sum_{n=2}^\infty \sum_{m=0}^n \frac{(n+1) P^m_n}{r^{n+2}} \times \nonumber \\ 
&&\bigg\{ \left[ \frac{A_{mn} r^2}{2(2n-1) \eta} - B_{mn} \right] \cos m\phi  \nonumber \\
&+&\left[ \frac{\tilde{A}_{mn} r^2}{2(2n-1) \eta} - \tilde{B}_{mn} \right]\sin m\phi   \bigg\}, 
\end{eqnarray}
\begin{eqnarray}
v_\theta =&& -\frac{1}{r^3} \bigg[ \left(B_{11}- \frac{a^2}{6 \eta} A_{11}\right) \cos\theta \cos\phi + \left(\tilde{B}_{11}-\frac{a^2}{6\eta} \tilde{A}_{11} \right) \cos\theta \sin\phi \nonumber \\
&+&  \left(B_{01}-\frac{a^2}{6\eta} A_{01}\right)\sin\theta   \bigg]   + \sum_{n=2}^\infty \sum_{m=0}^n \frac{\sin \theta P_n^{m'}}{r^{n+2}} \times \nonumber \\
&&\left\{ \left[ \frac{(n-2)r^2}{2n (2n-1) \eta} A_{mn} - B_{mn}\right] \cos m \phi + \left[ \frac{(n-2)r^2}{2n(2n-1) \eta} \tilde{A}_{mn} -  \tilde{B}_{mn} \right] \sin m\phi  \right\} \nonumber\\
& +& \sum_{n=2}^\infty \sum_{m=0}^n  \frac{mP^m_n}{r^{n+1}\sin \theta} \left(\tilde{C}_{mn} \cos m\phi - C_{mn} \sin m\phi \right), 
\end{eqnarray}
\begin{eqnarray}
v_\phi =&&   \frac{1}{r^3} \left[ \left(B_{11}-\frac{a^2}{6 \eta}A_{11} \right)\sin \phi  - \left(\tilde{B}_{11}-\frac{a^2}{6 \eta}\tilde{A}_{11} \right) \cos\phi  \right]  \nonumber \\
&+& \sum_{n=2}^\infty \sum_{m=0}^n \frac{\sin \theta P_n^{m'}}{r^{n+1}} \left( C_{mn} \cos m\phi + \tilde{C}_{mn} \sin m\phi \right) \nonumber \\
&-& \sum_{n=2}^\infty \sum_{m=0}^n \frac{mP_n^m}{r^{n+2}\sin \theta} \bigg\{ \left[ \frac{(n-2)r^2}{2n(2n-1)\eta} \tilde{A}_{mn} -  \tilde{B}_{mn}  \right] \cos m\phi \nonumber \\
&-&\left[ \frac{(n-2)r^2}{2n(2n-1)\eta} A_{mn} - B_{mn}  \right] \sin m\phi   \bigg\} .
\end{eqnarray}

\section{Rate of work with radial deformation}\label{app:AppendixWork}
We now compute the rate of working of the swimmer with radial deformation. Lengthy calculations allow to compute the integral  from Eq.~\ref{eqn:P} as
\begin{eqnarray}
\mathcal{P}=&& \frac{48 \pi \eta}{a^5} \left( B_{01}^1 + B_{11}^2 + \tilde{B}_{11}^2  \right) + \frac{4 \pi}{3 a \eta} \left( A_{01}^2 + A_{11}^2 + \tilde{A}_{11}^2 \right) \nonumber \\
&-& \frac{16 \pi}{a^3} \left( A_{01}B_{01}+A_{11} B_{11}+\tilde{A}_{11} \tilde{B}_{11} \right) \nonumber \\
&+& \sum_{n=2}^\infty \frac{4 \pi n (n+1)}{2n+1} \bigg[ \frac{2n^3 + n^2 -2n+2}{2n^2 (2n-1)^2 \eta a^{2n-1}} A_{0n}^2 - \frac{2(n+2)}{n a^{2n+1}} A_{0n}B_{0n} \nonumber \\
&+& \frac{(10n+4 +4n^2) \eta}{n a^{2n+3}} B_{0n}^2 + \frac{(n+2) \eta}{a^{2n+1}} C_{0n}^2  \bigg] \nonumber \\
&+& \sum_{n=2}^\infty \sum_{m=1}^n \frac{2\pi n (n+1)(n+m)!}{(2n+1)(n-m)!} \Bigg[ \frac{2n^3+n^2-2n+2}{2n^2 (2n-1)^2 \eta a^{2n-1}} (A_{mn}^2+ \tilde{A}_{mn}^2) \nonumber \\
&-& \frac{2(n+2)}{n} (A_{mn} B_{mn}+\tilde{A}_{mn} \tilde{B}_{mn})  +\frac{(10n +4+4n^2) \eta}{n a^{2n+3}} (B_{mn}^2+ \tilde{B}_{mn}^2) \nonumber \\
&+& \frac{(n+2) \eta}{a^{2n+1}} (C_{mn}^2 + \tilde{C}_{mn}^2) \Bigg].
\end{eqnarray}
We have again employed Eq.~\ref{eqn:Useful} to obtain the above result. Removing all the non-axisymmetric modes and transforming the coefficients with Eqs.~\ref{eqn:PakLB1}--\ref{eqn:PakLB2}, the above expression agrees with the results in Blake \cite{Blake1971b}. For the case of purely tangential deformation, with Eq.~\ref{eqn:purelyTangential}, the above expression reduces to Eq.~\ref{eqn:PowerTangential} in the main text.

\section{Squirming with arbitrary surface velocities and radial deformation}\label{app:AppendixGeneral}
Here we allow in  the general squirming profile additional radial velocity components of the form $D_{m} (\theta), \tilde{D}_m (\theta)$,
\begin{eqnarray}
u_r|_{r=a}= \sum_{m=0}^\infty D_m (\theta) \cos m\phi + \tilde{D}_m (\theta) \sin m \phi, \label{eqn:project1}\\
u_\theta|_{t=a} =\sum_{m=0}^\infty E_m (\theta) \cos m \phi + \tilde{E}_m (\theta) \sin m \phi, \label{eqn:project2}\\
u_\phi|_{r=a}=\sum_{m=0}^\infty F_m (\theta) \cos m \phi + \tilde{F}_m (\theta) \sin m \phi, \label{eqn:project3}
\end{eqnarray}
and follow the same matching conditions, Eqs.~\ref{eqn:matching1}--\ref{eqn:matching2}, in order to determine the coefficients $A_{mn}$, $\tilde{A}_{mn}$, $B_{mn}$, $\tilde{B}_{mn}$, $C_{mn}$, $\tilde{C}_{mn}$ in Lamb's general solution. The only difference is that we no longer have the purely tangential deformation condition (Eq.~\ref{eqn:purelyTangential}), and therefore are required to determine $A_{mn}, \tilde{A}_{mn},B_{mn}, \tilde{B}_{mn}$ such that the radial (Eq.~\ref{eqn:matching1}) and polar (Eq.~\ref{eqn:matching2}) matching conditions are satisfied simultaneously. Using the orthogonality of the associated Legendre polynomials and  simultaneously solving the equations, we obtain
\begin{eqnarray}
A_{mn} &=& \frac{(2n-1) a^n \eta}{2} \frac{(2n+1)(n-m)!}{(n+1)(n+m)!} \times \nonumber \\
&&\int_{-1}^1 \left[ n D_m + \frac{\partial}{\partial \mu} \left(E_m \sin \theta \right)- \frac{m \tilde{F}_m}{\sin \theta} \right] P_n^m d\mu ,\\
\tilde{A}_{mn} &=& \frac{(2n-1) a^n \eta}{2} \frac{(2n+1)(n-m)!}{(n+1)(n+m)!} \times \nonumber \\
&&\int_{-1}^1 \left[ n \tilde{D}_m + \frac{\partial}{\partial \mu} \left( \tilde{E}_m \sin \theta\right)+ \frac{m F_m}{\sin \theta} \right] P_n^m d\mu, \\
B_{mn} &=& \frac{a^{n+2}}{4} \frac{(2n+1)(n-m)!}{(n+1)(n+m)!} \times \nonumber \\
&& \int_{-1}^1 \left[ (n-2) D_m + \frac{\partial}{\partial \mu} \left(E_m \sin \theta \right) -\frac{m \tilde{F}_m}{\sin \theta} \right] P_n^m d\mu, \\
\tilde{B}_{mn} &=& \frac{a^{n+2}}{4} \frac{(2n+1)(n-m)!}{(n+1)(n+m)!} \times \nonumber \\
&&\int_{-1}^1  \left[ (n-2) \tilde{D}_m + \frac{\partial}{\partial \mu} \left( \tilde{E}_m \sin \theta\right)+ \frac{m F_m}{\sin \theta} \right] P_n^m d\mu,\\
C_{mn} &=& \frac{a^{n+1}}{2n}\frac{(2n+1)(n-m)!}{(n+1)(n+m)!}  \times \nonumber \\
&&\int_{-1}^1 \left[ -\frac{\partial}{\partial \mu} \left( F_m \sin \theta \right) -\frac{m \tilde{E}_m}{\sin \theta} \right] P_n^m d\mu, \\
\tilde{C}_{mn} &=& \frac{a^{n+1}}{2n}\frac{(2n+1)(n-m)!}{(n+1)(n+m)!}  \times \nonumber \\
&&\int_{-1}^1 \left[ -\frac{\partial}{\partial \mu} \left( \tilde{F}_m \sin \theta \right) +\frac{m E_m}{\sin \theta} \right] P_n^m d\mu,
\end{eqnarray}
for $0\le m \le \infty$ and $m \le n \neq0$. When there is no radial deformation, i.e.~$D_m = \tilde{D}_m= 0$, we recover the results without radial deformation (Eqs.~\ref{eqn:ProjectResultsTan1}--\ref{eqn:ProjectResultsTan2} \& \ref{eqn:purelyTangential}).

\end{appendix}

\end{article}
\end{document}